\documentclass[twocolumn]{aa}

  \usepackage{amscd}
  \usepackage{amsmath}
  \usepackage{amssymb}
  \usepackage[T1]{fontenc}
  \usepackage[varg]{txfonts}
  \usepackage{graphicx}
  \usepackage{natbib}
  \usepackage{verbatim}
  \usepackage{array}
  \usepackage{aalongtable}

\begin{document}

\title{Type Ia supernova diversity in three-dimensional models}

\authorrunning{F.~K.~R{\"o}pke et al.}

\author{F. K. R{\"o}pke\inst{1},  M. Gieseler\inst{1},
        M. Reinecke\inst{1}, C. Travaglio\inst{2}
        \and
        W. Hillebrandt\inst{1}}
          
   \offprints{F. K. R{\"o}pke}

   \institute{Max-Planck-Institut f\"ur Astrophysik,
              Karl-Schwarzschild-Str. 1, D-85741 Garching, Germany\\
              \email{[fritz;mccg;martin;wfh]@mpa-garching.mpg.de}
         \and
              INAF -- Osservatorio Astronomico di Torino,
              Strada dell'Osservatorio 20, I-10025 Pino Torinese,
              Torino, Italy\\
              \email{travaglio@to.astro.it}
             }

\abstract{
The use of type Ia supernovae as distance indicators for cosmology has
initiated a search for theoretical arguments supporting the empirical
calibration methods applied. To this end, as a first step, a sound
understanding of the origin of the observed diversity in type Ia
supernova properties is needed. Here we present a first systematic
study of effects resulting from changing some physical parameters of
three-dimensional deflagration models of thermonuclear supernovae. In our
study we vary the progenitor's carbon-to-oxygen ratio and its central
density prior to ignition because both properties are not well
determined by stellar evolution theory and they may change from
supernova to supernova. Next we compute for these explosion models the
nucleosynthesis yields in a post-processing step. This, in addition,
allows us to study variations in the progenitor's metallicity by means
of different $^{22}$Ne mass fractions in the initial composition. We
find that the progenitor's carbon-to-oxygen ratio and its central
density affect the energy release of the models and thus the expansion
velocity of the supernova. Moreover, we find that changing the
metallicity and the central density changes the production of
radioactive $^{56}$Ni and thus affects the luminosity. In contrast,
the carbon-to-oxygen ratio has little effect on the $^{56}$Ni
production. Implications of the found variations of the explosion
energy and the produced $^{56}$Ni mass for the type Ia supernova
diversity are discussed.

} 

\maketitle

\section{Introduction}
\label{intro}

Type Ia supernovae (SNe Ia) have become a major tool to determine
cosmological parameters. As a consequence of their uniformity of
properties and their enormous brightness they are suitable to be 
applied in cosmological distance measurements. 
However, they cannot be
claimed to be perfect ``standard candles'', as they show a significant
intrinsic scatter in their peak luminosities as well as other
characteristics. Therefore their cosmological application rests on
empirical corrections of the peak luminosities based on correlations
with other observables \citep[e.g.][]{phillips1993a}. Only such empirical
corrections facilitated
distance measurements of SNe Ia at high redshifts which have led to
the spectacular conclusion
that the expansion of universe currently accelerates
\citep{riess1998a,perlmutter1999a}. One possibility to technically
take this result into account is by a cosmological constant in the
Einstein equations eventually indicating a dark energy component of
the universe (for a review see \citealt{leibundgut2001a}).

This underscores the need for a theoretical reasoning of the
correlation of characteristics that are yet established only
empirically. A theoretical understanding will help to answer questions,
such as a possible affliction of calibration procedures with evolutionary
effects. 

In the astrophysical standard model \citep[see][]{hillebrandt2000a},
SNe Ia are associated with thermonuclear explosions of carbon/oxygen
white dwarf (WD) stars. The optical event is powered by the decay of
radioactive species (e.g.\  $^{56}$Ni) produced in the thermonuclear
burning. 
Numerical simulations on the basis of this scenario provide an
approach to the understanding of
calibration methods. Recently,
there has been much progress in the three-dimensional modeling of the
explosion process \citep{hillebrandt2000b,reinecke2002b,reinecke2002c,
reinecke2002d,gamezo2003a} and the question arises whether it is
possible to reproduce the SN Ia diversity by varying the initial
parameters of such models. This will be addressed in the present study where
we restrict the survey to so-called deflagration models of
thermonuclear supernovae which can be summarized as follows.

After
ignition near the center of the WD the flame propagates outward
in the subsonic deflagration mode,  i.e.\ it is mediated by
thermal conduction of the degenerate electron gas. This outward
burning produces an inverse density stratification in the gravitational
field of the WD star with dense fuel on top of hot and light
ashes. Consequently, due to buoyancy (Rayleigh-Taylor) instabilities
burning bubbles form that rise into the fuel leading to shear
flows.
Kelvin-Helmholtz instabilities 
generate strong turbulence given the fact that the Reynolds number
typical for the situation is of the order of $10^{14}$.
Resulting from this, turbulent eddies decay to smaller scales
thereby forming a turbulent energy cascade. The interaction of the
thermonuclear flame with these turbulent motions is the key feature of
the deflagration model for SNe Ia. A laminar flame would burn far too
slowly to release sufficient energy for an explosion of the
star. However, 
the wrinkling of the flame due to turbulence and the accompanying
flame surface enhancement increase the net burning rate and accelerate
the flame. This defines the deflagration model of thermonuclear
supernova explosions as a problem of turbulent
combustion. \citet{reinecke2002d} and \citet{gamezo2003a} could show
that this model indeed leads to an explosion. Whether it reproduces
all aspects of observed supernovae is still not fully explored
\citep[e.g.][]{kozma2005a}. \citet{gamezo2004a} and
\citet{hoeflich1998a} claim that a 
hypothetical transition from the deflagration mode of flame
propagation to a supersonic detonation needs to be invoked at later
phases of the explosion. We set aside such a transition because its
physical origin in not understood \citep{niemeyer1999a}. Moreover,
even in
such a case the initial deflagration stage will be essential for
understanding the SN Ia diversity since large fractions of the energy
and the radioactive $^{56}$Ni (which powers the lightcurve) are
produced here and nonlinear effects in flame propagation are extremely
sensitive to the initial conditions. 

The crucial role played by three-dimensional effects in deflagration
SN Ia models calls for multi-dimensional simulations to study the
diversity of such events.
Most previous attempts to unveil the origin of the SN Ia
diversity were, however, based on one-dimensional models
\citep{bravo1993a, bravo1996a,
hoeflich1998a, umeda1999b, iwamoto1999a, dominguez2000a,
dominguez2000b, dominguez2001a}. These are hampered by introducing
free parameters due to
incomplete description of the relevant physics in addition to the
initial parameters they intend to study. The description of the
turbulent mixing as well as the effective flame velocity is not
inherent in one-dimensional models but rather accomplished
in a parametrized way. Due to the free parameters empirical
one-dimensional models are not 
sufficiently predictive to nail down explanations for the diversity of
SNe Ia, but they can nevertheless provide valuable clues for possible
trends. 
Systematic studies based on three-dimensional models overcome the
ambiguity of the turbulent flame velocity and mixing. By correctly
modeling these effects, multi-dimensional deflagration models contain
no tunable parameters and possess a high predictive power.
However, due to the challenging computational demands of
three-dimensional models, the available
studies of initial parameters are very incomplete.

Applying a simplified setup we present the first systematic
survey of the impact of initial parameters on three-dimensional SN Ia
models. The price of the simplicity (and possibly incompleteness) of
our models is that we cannot set the absolute scale of the effects in
the presented parameter study. Nevertheless, we are able to point out
the trends of effects from varying the initial parameters.

We restrict this first systematic study to variations of
the central density, the initial carbon-to-oxygen (C/O) ratio and the
metallicity of the progenitor just prior to ignition.
Our intention is
to test the parameters independently, setting aside any realistic
evolution of the progenitor system. For detailed progenitor evolution
studies we refer to e.g.\
\citet{nomoto1985a}, \citet{hernanz1988a}, \citet{bravo1996a},
\citet{umeda1999a}, \citet{langer2000a}, and \citet{dominguez2001a}.
Important parameters that are not addressed in this study are for
instance rotation and the way of flame
ignition \citep[see e.g.][]{woosley2004a}. Some effects of the
ignition conditions on SN Ia models and nucleosynthesis
have been recently discussed by \citet{travaglio2004a}.

In Sect.~\ref{num_model} we describe the numerical schemes we apply to
model SNe Ia explosions and the nucleosynthesis, followed by a discussion
of the parameter space to be explored in Sect.~\ref{param}. The
features of the explosion models will be compared in Sect.~\ref{expl},
and Sect.~\ref{nuc} describes the nucleosynthetic yields of these
models. Conclusions are drawn in Sect.~\ref{concl}

\section{Numerical Model}

The numerical model applied in our study consists of two parts. In a
first step we simulate the hydrodynamics of the explosion
process. Here, the description of the nuclear processes is very
coarse. With the information gained from tracer particles advected in
this simulation we perform a nucleosynthetic postprocessing as a
second step. This enables us to infer the production of the individual
isotopes. Both methods will be briefly described in the following.

\label{num_model}
\subsection{Explosion dynamics}
\label{exp_model_sect}

The deflagration model of thermonuclear supernova explosions as
outlined in Sect.~\ref{intro} was
implemented in a numerical scheme by
\citet{reinecke1999b,reinecke2002b}. We refer to these works for the
details of the applied techniques and will only mention the basic
aspects here. 

The major problem of SN Ia simulations is the vast range of relevant
scales. The thickness of the flame is tiny compared with the
dimensions of the WD star and the turbulent cascade interacts with the
flame down to the so-called Gibson scale where the turbulent velocity
fluctuations become comparable with the laminar flame speed.
Neither the internal flame structure nor the Gibson scale can be
resolved in multidimensional simulations in the foreseeable future and
thus the flame propagation and turbulence effects have to be
adequately modeled in numerical simulations.

Seen from  the size of the WD star, it is well justified to
regard the unresolved flame as a discontinuity separating the fuel from
the ashes. Then the description of flame propagation has to track this
interface and a technique well-suited for this purpose is the
so-called \emph{level set method} \citep{osher1988a}. It is widely
used in simulations of combustion problems in engineering. In this
technique, the flame front is associated with the zero level set of a
scalar field $G$. For numerical reasons, $G$ is chosen to be a signed
distance function with respect to the flame front. To model the flame
propagation we evolve the
$G$-field according to the scheme described by
\citet{reinecke1999a}.

In this scheme the effective flame velocity has
to be provided. To this end, the notion  is essential that turbulent
combustion proceeds in different regimes
\citep[e.g.][]{peters2000a}. For most parts of the supernova explosion
the so-called \emph{flamelet regime}
applies, where the flame as a whole is wrinkled by turbulence. Here,
the flame propagation is known to decouple from the microphysics of
the burning process and to be determined by the turbulent motions
exclusively \citep{damkoehler1940a}. 
These, however, are derived from a \emph{subgrid scale model}
implemented first in SN Ia simulations by \citet{niemeyer1995b}. It
describes the effects turbulence on unresolved scales. In this sense
our model can be
regarded as a Large Eddy Simulation (LES) well-known from
computational fluid dynamics.
Since flame propagation is modeled in our simulations, supplementary
simulations of the physical processes on small scales have to be
provided that ensure the validity of the underlying
assumptions. In this spirit \citet{roepke2003a, roepke2004a,
  roepke2004b} showed that flame
propagation proceeds in a stabilized way below the Gibson scale.

The hydrodynamics is modeled based on the PROMETHEUS implementation
\citep{fryxell1989a} of the piecewise parabolic method
\citep{colella1984a}. The equation of state of the WD material
comprises contributions from a variably degenerate and
relativistic electron gas, ions following the Boltzmann statistics, a
photon gas and eventually electron/positron pairs.

The correct way to incorporate the nuclear burning would require a
full reaction network. However, due to the restricted computational
resources only a very simplified description of the nucleosynthesis is
possible concurrent with the explosion simulation. Our implementation
follows the approach suggested by \citet{reinecke2002b}, who include
five species, viz.\ $\alpha$-particles, $^{12}$C, $^{16}$O, ``Mg'' as
a representative of intermediate mass elements and ``Ni'' representing
iron group nuclei\footnote{In the following we set ``Ni'' and ``Mg''
  in quotes when we refer to the iron group elements and intermediate
  mass elements followed in the explosion hydro-simulations. This is
  done to avoid confusion with the results of the nuclear
  postprocessing.}. The fuel is assumed to be a mixture of carbon and
oxygen. At the initial high densities burning proceeds to nuclear
statistical equilibrium (NSE) composed of $\alpha$-particles and
``Ni''. Depending on temperature and density in the ashes, the NSE
composition changes, which has significant impact on the explosion
dynamics \citep{reinecke2002b}. Once the fuel density drops below
$5.25 \times 10^7 \,\mathrm{g}\,\mathrm{cm}^{-3}$ due to the expansion
of the WD, burning is assumed to terminate at intermediate mass
elements. Below $1 \times 10^7 \,\mathrm{g}\,\mathrm{cm}^{-3}$ burning
is switched off, since it is then expected to leave the flamelet
regime and to enter the so-called distributed burning regime. Here
turbulence penetrates the internal structure of the flame. This effect
is ignored in the present study but was addressed by
\citet{roepke2005a}.

In order to achieve a more detailed analysis of the nucleosynthetic
yields of the simulated supernova explosion we advect tracer particles
with the fluid motions recording temperature, density, and internal
energy as a function of time. These data then serve as input for a
nucleosynthetic postprocessing.

\subsection{Nuclear postprocessing}

The nuclear postprocessing determines the nucleosynthetic
yields of the explosion models \emph{a posteriori} from the data
recorded by the tracer particles. The applied method is similar to
that described by \citet{thielemann1986a} (there labeled as
\emph{method (a) simple postprocessing)}. Its application to SNe Ia
explosions is discussed in detail by \citet{travaglio2004a}.

\begin{table}
\centering
\caption{Nuclear reaction network (note that the elements below
  arsenic are irrelevant for SNe Ia).
\label{network_tab}}
\setlength{\extrarowheight}{2pt}
\begin{tabular}{p{0.15\linewidth}p{0.25\linewidth}|p{0.15\linewidth}p{0.25\linewidth}}
\hline\hline
element &  atomic mass $A$ & element &  atomic mass $A$\\
\hline
n  & 1              & Sc & 40 \ldots 50 \\
p  & 1              & Ti & 42 \ldots 52 \\
He & 4, 6           & V  & 44 \ldots 54 \\
Li & 6, 7, 8        & Cr & 46 \ldots 56 \\
Be & 7, 9, 10, 11   & Mn & 48 \ldots 58 \\
B  & 8, 9 \ldots 12 & Fe & 50 \ldots 62 \\
C  & 10 \ldots 15   & Co & 52 \ldots 63 \\
N  & 12 \ldots 17   & Ni & 54 \ldots 67 \\
O  & 14 \ldots 20   & Cu & 56 \ldots 69 \\
F  & 17 \ldots 21   & Zn & 59 \ldots 72 \\
Ne & 18 \ldots 25   & Ga & 61 \ldots 76 \\
Na & 20 \ldots 26   & Ge & 63 \ldots 78 \\
Mg & 21 \ldots 28   & As & 71 \ldots 80 \\
Al & 23 \ldots 30   & Se & 74 \ldots 83 \\
Si & 25 \ldots 33   & Br & 75 \ldots 83 \\
P  & 27 \ldots 35   & Kr & 78 \ldots 87 \\
S  & 29 \ldots 38   & Rb & 79 \ldots 87 \\
Cl & 31 \ldots 40   & Sr & 84 \ldots 91 \\
Ar & 33 \ldots 44   & Y  & 85 \ldots 91 \\
K  & 35 \ldots 46   & Nb & 91 \ldots 97 \\
Ca & 37 \ldots 49   & Mo & 92 \ldots 98 \\
\hline
\end{tabular}
\end{table}

The employed nuclear reaction network code was kindly provided by
F.-K.~Thielemann. It comprises 384 isotopes which are listed in
Table~\ref{network_tab} and takes into account $\beta$-decays,
electron captures, photo-disintegrations, two-body reactions, and
three-body reactions. A detailed description of the network is given
by \citet{thielemann1996a} and \citet{iwamoto1999a}. As
\citet{brachwitz2000a} and \citet{thielemann2003a} discussed
previously, the new electron capture and $\beta$-decay rates by
\citet{langanke2000a} and \citet{martinez-pinedo2000a} are included in
the network.

Since the description of the nuclear reactions in the hydrodynamic
explosion simulation is coarse and $Y_e$ is assumed to be constant at
a value of 0.5,
the internal energy recorded by the tracer particles is employed to
calculate a realistic temperature form a high-temperature equation of
state \citep{timmes2000a} combined with an improved nuclear reaction
network \citep[cf.][]{travaglio2004a}.

The nucleosynthesis is calculated separately for each
tracer particle. To level out variations in the data from the
hydrodynamic simulation, the minimal temperature is set to $10^9 \,
\mathrm{K}$. This measure guarantees stability of the nuclear reaction
network code. Subsequently, the maximum temperature $T_\mathrm{max}$ is
checked. If it does not exceed $2 \times 10^9 \, \mathrm{K}$, then the
corresponding material is treated as unprocessed.  This approach is
justified since the fuel consists only of $^{12}$C, $^{16}$O, and
$^{22}$Ne, which below $2\times 10^9 \, \mathrm{K}$ will burn hydrostatically,
not significantly contributing to the nucleosynthetic yields over the
simulated period of time.

For tracers with $T_\mathrm{max} > 2 \times 10^9 \, \mathrm{K}$ the
following procedure is applied:

\begin{enumerate}
\item \label{step1}
Nuclear statistical equilibrium (NSE) is assumed if the temperature of
the current time step $t_i$ is larger than $6 \times 10^9 \, \mathrm{K}$,
i.e.\ the strong reactions can be neglected and only the ``weak''
nuclear network is applied updating $Y_e$. Otherwise the full reaction network is
employed. 
\item \label{step2}
Temperature and density are interpolated for the sample point at
$t_{i+1}$. If these variables change for more than 5\% in the interval
$[t_i, t_{i+1}]$, the time step is halved.
\item
The network is solved for $t_{i+1}$. If a relative accuracy of
$10^{-5}$ cannot be reached in a limited number of steps, the time
step is again halved and we resume with point \ref{step2} of the
scheme. If 
this measure fails, the tracer is ignored in the final
result. Fortunately the number of such cases could be drastically
reduced to at most one out of $[27]^3$. When reaching NSE the new
abundances are calculated for the updated $Y_e$ at
$t_{i+1}$.
\item
If the abundance of an isotope drops below $10^{-25}$, it is set to
zero.
\end{enumerate}

\section{Parameter space}
\label{param}

The initial parameters we explore in our study 
(the carbon mass fraction $X(^{12}\mathrm{C})$, the  central density
$\rho_c$, and the metallicity $Z$ of the WD at ignition) are treated
as independent. This allows to disentangle the effects of the
individial parameters on the explosion process. Nontheless, the
parameter space is chosen in agreement with values suggested by
stellar evolution, as described below.

Different values for the central density of the
WD and the carbon-to-oxygen ratio of its material are applied in the
explosion model itself. Contrary to that, we vary the metallicity only
in the nucleosynthesis postprocessing. The nomenclature of the models
is given in Table~\ref{models_tab}.

\begin{table}
\centering
\caption{Model parameters.
\label{models_tab}}
\setlength{\extrarowheight}{2pt}
\begin{tabular}{p{0.15\linewidth}p{0.2\linewidth}p{0.15\linewidth}p{0.2\linewidth}l}
\hline\hline
model & $\rho_c$
 [$10^9\,\mathrm{g}\,\mathrm{cm}^{-3}$] & $X(^{12}\mathrm{C})$ &
 metallicity \\
\hline
\emph{1\_1\_1} & 1.0 & 0.30 & $0.5 Z_\odot$\\
\emph{1\_1\_2} & 1.0 & 0.30 & $1.0 Z_\odot$\\
\emph{1\_1\_3} & 1.0 & 0.30 & $3.0 Z_\odot$\\
\hline
\emph{1\_2\_1} & 1.0 & 0.46 & $0.5 Z_\odot$\\
\emph{1\_2\_2} & 1.0 & 0.46 & $1.0 Z_\odot$\\
\emph{1\_2\_3} & 1.0 & 0.46 & $3.0 Z_\odot$\\
\hline
\emph{1\_3\_1} & 1.0 & 0.62 & $0.5 Z_\odot$\\
\emph{1\_3\_2} & 1.0 & 0.62 & $1.0 Z_\odot$\\
\emph{1\_3\_3} & 1.0 & 0.62 & $3.0 Z_\odot$\\
\hline
\emph{2\_1\_1} & 2.6 & 0.30 & $0.5 Z_\odot$\\
\emph{2\_1\_2} & 2.6 & 0.30 & $1.0 Z_\odot$\\
\emph{2\_1\_3} & 2.6 & 0.30 & $3.0 Z_\odot$\\
\hline
\emph{2\_2\_1} & 2.6 & 0.46 & $0.5 Z_\odot$\\
\emph{2\_2\_2} & 2.6 & 0.46 & $1.0 Z_\odot$\\
\emph{2\_2\_3} & 2.6 & 0.46 & $3.0 Z_\odot$\\
\hline
\emph{2\_3\_1} & 2.6 & 0.62 & $0.5 Z_\odot$\\
\emph{2\_3\_2} & 2.6 & 0.62 & $1.0 Z_\odot$\\
\emph{2\_3\_3} & 2.6 & 0.62 & $3.0 Z_\odot$\\
\hline
\end{tabular}
\end{table}

\subsection{Variation of the carbon mass fraction}

The origin of the diversity in the carbon mass fraction has been
studied by \citet{umeda1999a} by numerically evolving the corresponding
binary systems with 3--9 $M_\odot$ WD progenitor stars. They found it
to depend on the metallicity and the zero-age main sequence (ZAMS)
mass of the WD progenitor, as well as on the mass of the companion
star. These in turn determine the mass of the WD, $M_{\mathrm{WD},0}$,
just prior to the onset of accretion. The main outcome of the survey was that
$X(^{12}\mathrm{C})$ in the core of the WD decreases with increasing
$M_{\mathrm{WD},0}$ and that the direct dependence of  $X(^{12}\mathrm{C})$
on the metallicity is small although the correlation between the ZAMS
mass and $M_{\mathrm{WD},0}$ depends sensitively on it
\citep{umeda1999b}. Taking into account the conditions ensuring that
the WD will accrete mass until reaching $M_\mathrm{Ch}$,
\citet{umeda1999a} infer that $X(^{12}\mathrm{C})$ may vary in the
range from $\sim$$0.36$ to $\sim$$0.5$. These values apply only to the
convective core of the WD. The accreted material is assumed to be
processed to a C/O ratio of $\sim$$1$, leading to a gradient of the
carbon mass fraction inside the WD. This effect will be ignored in our
model, were we postulate a uniform C/O ratio throughout the entire
star employing values of 0.30, 0.46, and 0.62 for $X(^{12}\mathrm{C})$
(cf.~Table~\ref{models_tab}).

\subsection{Variation of the central density}

The variation of the central density in SN Ia progenitors just before
the ignition of the flame is even more difficult to constrain. At
least two effects determine the value of $\rho_c$. 

The first is the accretion history of the binary system \citep[see][for a detailed  study of the accretion
process]{langer2000a}. There seems to be only a narrow window in the range of
possible accretion rates $\dot{M}$ in which carbon can be ignited
centrally, avoiding off-center ignitions and gravitational collapse
due to high electron-capture rates. \citet{nomoto1985a} report on two
centrally ignited models with $\rho_c = 1.7 \times 10^9 \, \mathrm{g}
\, \mathrm{cm}^{-3}$ and $\rho_c = 5.2 \times 10^9 \, \mathrm{g} \,
\mathrm{cm}^{-3}$, respectively, and \citet{bravo1996a} find models in
the range $1.8 \times 10^9 \, \mathrm{g} \, \mathrm{cm}^{-3} \lesssim
\rho_c \lesssim 6.3 \times 10^9 \, \mathrm{g}\,
\mathrm{cm}^{-3}$. However, the exact range of that window is
uncertain and depends additionally on the white dwarf mass and
temperature. Initially cooler WDs are shifted to rather high central
densities in the range $6 \times 10^9 \, \mathrm{g} \,
\mathrm{cm}^{-3} \lesssim \rho_c  \lesssim 1.3 \times 10^{10} \,
\mathrm{g}\, \mathrm{cm}^{-3}$ \citep{hernanz1988a}.

The second effect is the establishment of the thermal structure of the
WD. Cooling due to plasmon neutrino losses and neutrino
bremsstrahlung have to be taken into account \citep{iwamoto1999a},
and a (most uncertain) contribution may come from the convective Urca
process \citep{paczynski1972a,barkat1990a,mochkovich1996a, lesaffre2005a}.

\citet{bravo1993a} calculate models for central densities at ignition
of $2.5 \times 10^9\,\mathrm{g} \,\mathrm{cm}^{-3}$, $4.0 \times
10^9\,\mathrm{g} \,\mathrm{cm}^{-3}$, and $8.0 \times 10^9
\,\mathrm{g} \,\mathrm{cm}^{-3}$, and  \citet{iwamoto1999a} use values
of $1.37 \times 10^9 \,\mathrm{g} \,\mathrm{cm}^{-3}$ and $2.12 \times
10^9 \,\mathrm{g} \,\mathrm{cm}^{-3}$. We assume central densities of
$1.0 \times 10^9 \,\mathrm{g} \,\mathrm{cm}^{-3}$ and $2.6 \times 10^9
\,\mathrm{g}
\,\mathrm{cm}^{-3}$ (see Table~\ref{models_tab}). Unfortunately it is
not yet possible to apply higher central densities, since then
electron captures will become dynamically important. Although electron
captures are correctly treated in the nuclear postprocessing, this effect is not
implemented in the current explosion models.

\subsection{Variation of the metallicity}

Our ignorance concerning realistic progenitor evolution is evident in
the approach to prescribe the metallicity $Z$ of the progenitor
independent of the other parameters. Detailed stellar evolution calculations
(e.g.~\citealt{umeda1999a,dominguez2001a}) have shown that it strongly
influences the progenitor's central density and also the C/O
ratio. Nevertheless, in the spirit of our exploration of possible
effects in the explosion models, we set aside a realistic progenitor
description and treat the metallicity as an independent parameter. 
A direct effect of the metallicity of the WD's progenitor is the
$^{14}$N abundance after the CNO burning phase. In Helium burning it
is then converted mostly to $^{22}$Ne. For simplicity, we
assume a uniform distribution of this $^{22}$Ne, which is only justified in
regions mixed by pre-ignition convection.

An analytic estimation of the effect of the metallicity on the
$^{56}$Ni production was given by \citet{timmes2003a}. They suggest a
variation of $Z$ ranging from $1/3$ to 3 times solar, based on
observations of field white dwarfs.
Following this suggestion we vary the $^{22}$Ne abundance in our
models to simulate a variation
in metallicity $Z$. In particular we explore $Z_\odot$
(corresponding to $X(^{22}\mathrm{Ne}) = 0.025$), $0.5 Z_\odot$, and
$3 Z_\odot$ (cf.~Table~\ref{models_tab}).

\section{Simulation setup}

\begin{figure*}[t]
\centerline{
\includegraphics[width = 0.72 \linewidth]
  {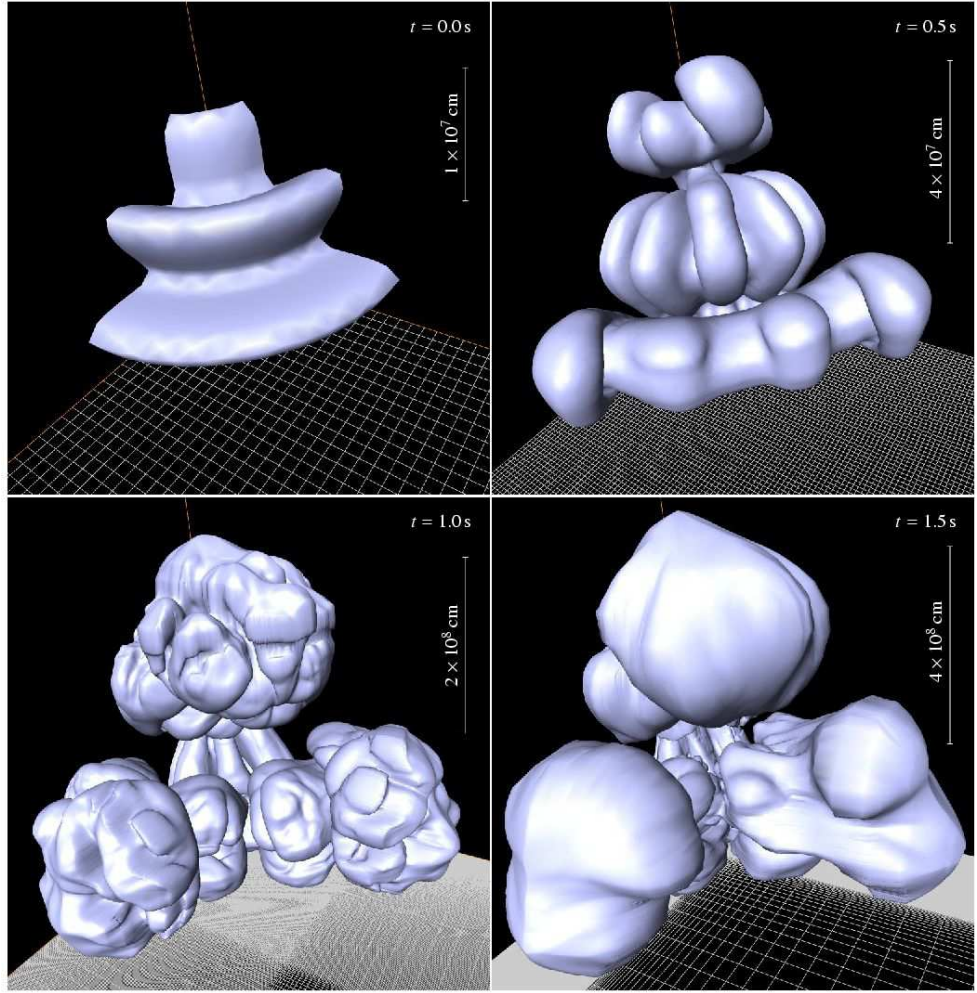}}
\caption{Time evolution of the burning front for model \emph{2\_2\_X}.
  \label{evo_fig}}
\end{figure*}

A rather large number of simulations is required to cover the
parameter space we aim at. We therefore have to minimize the
computational expenses by applying a simple setup for the individual
models. Our calculations span only one spatial octant and assume mirror
symmetry to the other octants. Full-star simulations
\citep{roepke2005b} have shown that this approach does not miss
large-scale flame features and thus -- although being
a simplification -- does not restrict the validity of the model. The
simulations were set up on a
Cartesian computational grid that was equally spaced in the inner
regions. To capture the expansion of the WD, the outer grid cells were
widened exponentially. Recently, \citet{roepke2004d} showed that with a
comoving computational grid the evolution can be followed to
homologous expansion. This, however, is not applied in the present
models.

The resolution of the individual runs was rather low -- the
computational domain was divided in $[256]^3$ grid cells corresponding
to a central grid resolution of $10^6 \, \mathrm{cm}$. In each
direction the grid length in the outer 35 zones was increased
subsequently by a factor of 1.15. As was pointed out by
\citet{reinecke2002c} the chosen resolution still guarantees the
explosion characteristics to be numerically converged (possibly with
the exception of the latest stages of the burning where intermediate
mass elements are produced). However, with this
resolution it is not possible to set up reasonable multi-point
ignition scenarios, since only a very small number of seed-bubbles
could be resolved. This is certainly a drawback because
\citet{reinecke2002d} showed that such models give rise to more
vigorous explosions. We restrict our simulations to
the centrally ignited model \emph{c3\_3d\_256} model of
\citet{reinecke2002c} in which the spherical initial flame geometry is
perturbed with three toroidal rings (see the upper left panel of
Fig.~\ref{evo_fig}).
Note that we initially incinerate the same volume in all models, which
does not correspond to the same mass for different central
densities. This ensures the same initial numerical resolution of the
flame front. 

For the construction of the WD near the Chandrasekhar mass we follow
the procedure described by \citet{reinecke_phd}. We assume a cold
isothermal WD of a temperature $T_0 = 5 \times 10^5 \,
\mathrm{K}$. With the chosen values for the carbon mass fraction of the
material and the central density we integrate the equations of
hydrostatic equilibrium using the equation of state described in
Sect.~\ref{exp_model_sect}. Depending on the central densities and
compositions the masses of the resulting WDs vary slightly: for $\rho_c
= 1.0 \times 10^9 \, \mathrm{g}\,\mathrm{cm}^{-3}$ and $\rho_c
= 2.6 \times 10^9 \, \mathrm{g}\,\mathrm{cm}^{-3}$ the WD masses amount
to $1.367\,M_\odot$ and $1.403\,M_\odot$, respectively.
% $\rho_c = 2.6 \times 10^9 \, \mathrm{g}\,\mathrm{cm}^{-3}$ : $1.416\,M_\odot$
As tested by \citet{reinecke_phd}, the construction procedure
guarantees stability of the WD over a time longer than simulated.

The $[n_\mathrm{trace}]^3$ tracer particles are distributed in an
$n_\mathrm{trace} \times n_\mathrm{trace} \times n_\mathrm{trace}$
equidistant grid in the integrated mass $M_0(r)$, the azimuthal angle
$\phi$, and $\cos \theta$, so that each particle represents the same
amount of mass. In order to improve the tracer particle statistics, a
random offset to the coordinates was applied. 
This offset was chosen small enough to keep the tracer
particles in their individual mass cells.
The values of the
density, the temperature and the internal energy at the tracer
particle's location and its coordinates were recorded every
$\sim$$1\,\mathrm{ms}$. This allows for an accurate reconstruction of
the trajectories as well as the final velocities and the
thermodynamical data. In the models presented in the following we set
$n_\mathrm{trace} = 27$. To test the representation of the model in
the tracer particles in cases of low central densities, this number was
increased to 35 in test calculations, as will be discussed below. 

\section{Explosion models}
\label{expl}

The explosion simulation for the exemplary case of model
\emph{2\_2\_X} (the metallicity does not affect the explosion dynamics
in our implementation) at four different times is illustrated in
Fig.~\ref{evo_fig}. The isosurface indicating the position of
the flame front is rendered from the zero level set of
the scalar field $G$. The computational grid plotted in these
snapshots visualizes our setup with
uniform grid cells in the inner region and an exponential growth of
the grid spacing further out. Our initial flame configuration is
shown in the upper left snapshot of Fig.~\ref{evo_fig}. In the
subsequent snapshots the growth of instabilities and an increasing
wrinkling of the flame front are visible. Once the flame enters the
exponentially growing part of the grid, the resolution of flame
features becomes coarser. However, at this stage the expansion of the
WD decreases the density of the fuel to values where
burning has largely ceased in our model. Thus the coarse flame
resolution in late stages of the simulation does not affect the
results.

\begin{figure}[t]
\centerline{
\includegraphics[width = \linewidth]
  {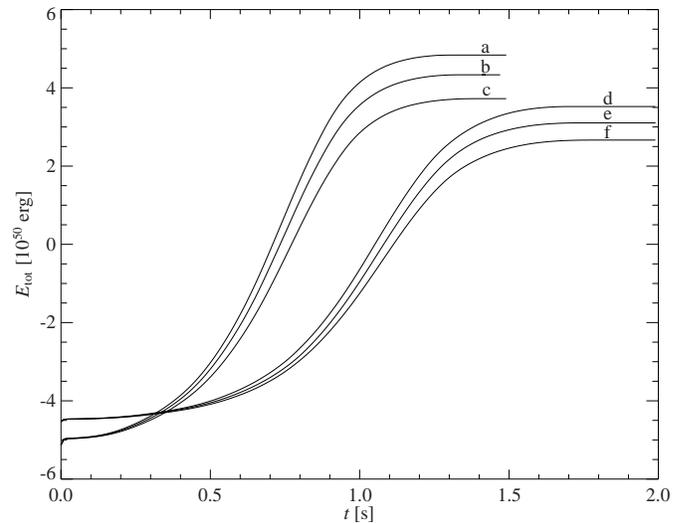}}
\caption{Total energies in models
  (a) \emph{2\_3\_X}, (b) \emph{2\_2\_X}, (c) \emph{2\_1\_X},  (d)
  \emph{1\_3\_X}, (e) \emph{1\_2\_X}, and (f) \emph{1\_1\_X}.
  \label{etot_fig}}
\end{figure}

\begin{table}
\centering
\caption{Results of explosion models: produced masses of iron group
  elements (``Ni'') and intermediate mass elements (``Mg''), nuclear
  energy release, and total energy at the end of the simulations.
\label{energy_tab}}
\setlength{\extrarowheight}{2pt}
\begin{tabular}{p{0.1\linewidth}p{0.15\linewidth}p{0.15\linewidth}
p{0.15\linewidth}p{0.15\linewidth}}
\hline\hline
model & $M(\mbox{``Ni''})$ $[M_\odot]$ & $M(\mbox{``Mg''})$ $[M_\odot]$
& $E_\mathrm{nuc}$ $[10^{50}\,\mathrm{erg}]$ & $E_\mathrm{tot}$ $[10^{50}\,\mathrm{erg}]$\\
\hline
\emph{1\_1\_X} & 0.3944 & 0.2067 & 6.974 & 2.714\\
\emph{1\_2\_X} & 0.3867 & 0.2081 & 7.445 & 3.140\\
\emph{1\_3\_X} & 0.3757 & 0.2144 & 7.870 & 3.563\\
\hline
\emph{2\_1\_X} & 0.5178 & 0.1874 & 8.851 & 3.772\\
\emph{2\_2\_X} & 0.5165 & 0.1859 & 9.461 & 4.412\\
\emph{2\_3\_X} & 0.5104 & 0.1822 & 9.966 & 4.909\\
\hline
\end{tabular}
\end{table}

\begin{figure}[t]
\centerline{
\includegraphics[width = \linewidth]
  {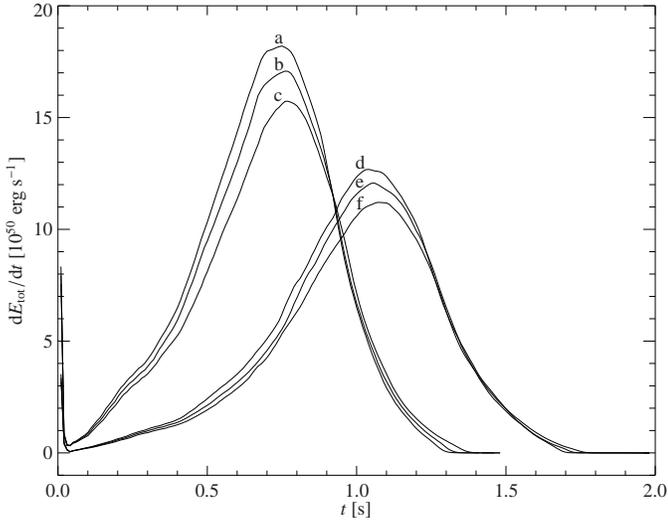}}
\caption{Energy generation rates in models
  (a) \emph{2\_3\_X}, (b) \emph{2\_2\_X}, (c) \emph{2\_1\_X}, (d)
  \emph{1\_3\_X}, (e) \emph{1\_2\_X}, and (f) \emph{1\_1\_X}.
  \label{egen_fig}}
\end{figure}

Fig.~\ref{etot_fig} shows the total energy production of our
models.
Due to the simple setup, all explosions are weak, but trends
can clearly be identified.  The energy releases of the different
models are listed in Table~\ref{energy_tab}, which also provides the
masses of produced iron group elements (``Ni'') and intermediate mass
elements (``Mg''). In Figs.~\ref{egen_fig}
and \ref{eturb_fig} the energy generation rates and
the evolution of the turbulent energies in our models are plotted,
respectively.

\subsection{Variation of the progenitor's C/O ratio}
\label{eplo_co}

The effects of a variation of the progenitor's carbon-to-oxygen ratio
on the SN Ia explosion models have been described by
\citet{roepke2004c}. We extend the discussion here.

Considering the explosion energetics first, Fig.~\ref{etot_fig} shows
that a higher carbon mass fraction leads to
an increased energy production for fixed central densities. Values are
given in Table~\ref{energy_tab}. For both central densities
the nuclear energy releases of the models increase by 12\% ($\sim$27\%
in the total energies) changing
$X(^{12}\mathrm{C})$ from 0.30 to 0.62.
The observed trend is not surprising and can easily be explained by the
burning process. The predominant effect is certainly the difference in
the mean binding
energy of the fuel.
A higher carbon mass fraction
increases the total energy
generation for the simple reason that the binding energy of
$^{12}\mathrm{C}$ is lower than that of $^{16}\mathrm{O}$ so that it
releases more energy by fusion to iron group elements.
A minor effect could be that the laminar burning velocity  increases
with $X(^{12}\mathrm{C})$ \citep{timmes1992a}.
This, however, is negligible in our models, since
already after a few time steps the flame propagation is completely
determined by the turbulent flame speed.

\begin{figure}[t]
\centerline{
\includegraphics[width = \linewidth]
  {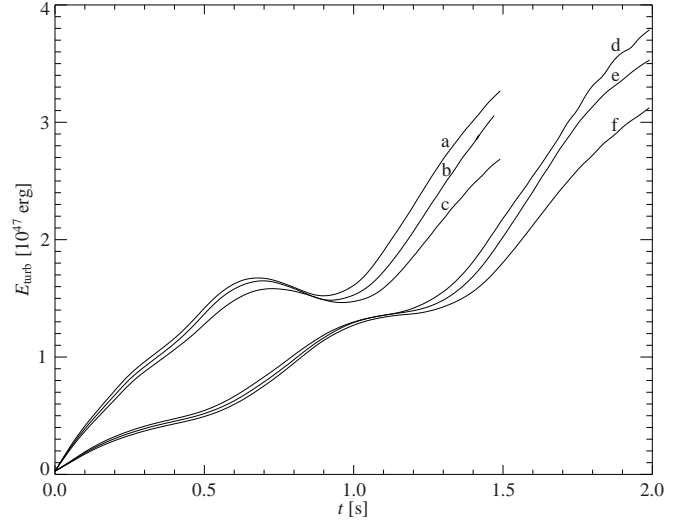}}
\caption{Turbulent energies in models
  (a) \emph{2\_3\_X}, (b) \emph{2\_2\_X}, (c) \emph{2\_1\_X}, (d)
  \emph{1\_3\_X}, (e) \emph{1\_2\_X}, and (f) \emph{1\_1\_X}.
  \label{eturb_fig}}
\end{figure}

\begin{figure*}[t]
\centerline{
\includegraphics[width = \linewidth]
  {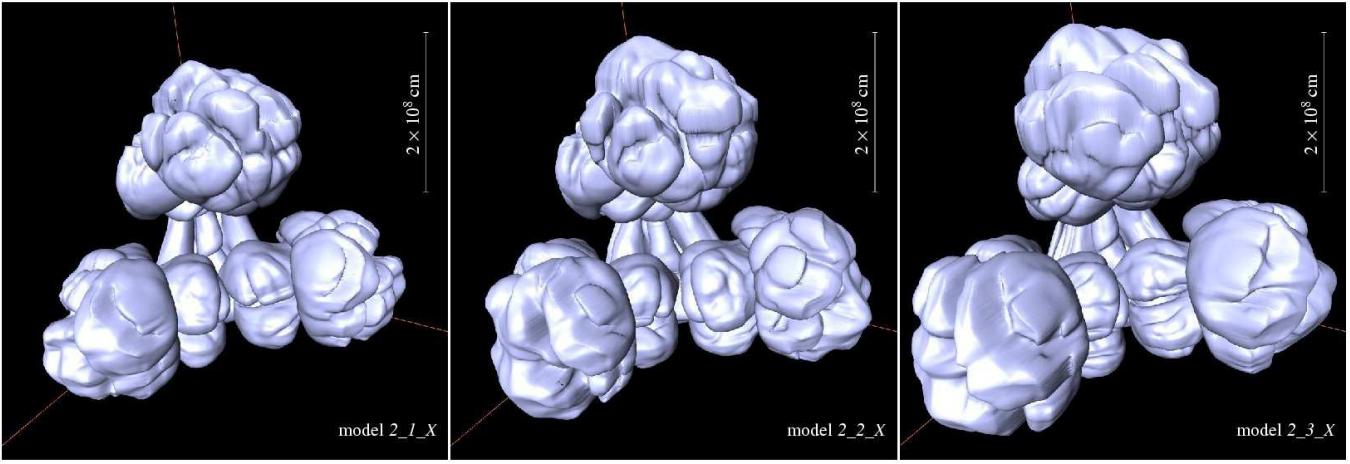}}
\caption{Flame surface of models with different carbon mass fraction
  at $t = 1.0 \, \mathrm{s}$.
  \label{grid_co_fig}}
\end{figure*}

It is noteworthy that the evolution of the energetics in the model
does not show a strong temporal shift. The energy generation rate
peaks at comparable times for the models with different carbon mass
fractions (cf.\ Fig.~\ref{egen_fig} for the temporal evolution of the
energy generation rate\footnote{Note, that the peak at
  $t=0\,\mathrm{s}$ is caused by our setup in which the initial flame
  is initialized by instantly incinerating the material behind it.}).
Fig.~\ref{etot_fig} reveals that the total energies of our models are
very similar for the largest part of energy generation and differ only
in the late phases. In this point our findings disagree with
\citet{khokhlov2000a}. Although he speculates that a decreasing
$X(^{12}\mathrm{C})$ would result in weaker explosions, he claims that
the reason is a delay of the development of the buoyancy
instabilities, which seems to be only a minor effect in our
simulations. The reason for the difference in the interpretation of the
results is possibly that the models of \citet{khokhlov2000a} were
apparently not evolved beyond the maximum of energy generation so
that it is difficult to distinguish between a delay and an overall
lower energy production.

Contrary to the explosion energetics, the produced masses of iron
group elements are unexpected.
The working hypothesis by \citet{umeda1999b} predicting a larger
production of $^{56}$Ni for larger carbon mass fractions was based on
the speculation that the resulting increased energy release would
enhance buoyancy effects and thus accelerate the turbulent flame
propagation. Consequently, more material would be burnt at high
velocities producing larger fractions of iron group elements. 
As emphasized by \citet{umeda1999b}, this hypothesis can only be
tested in multi-dimensional simulations which treat the turbulent
flame velocity in an unparametrized way. 
This is provided by our approach, but
\begin{figure}[t]
\centerline{
\includegraphics[width = \linewidth]
  {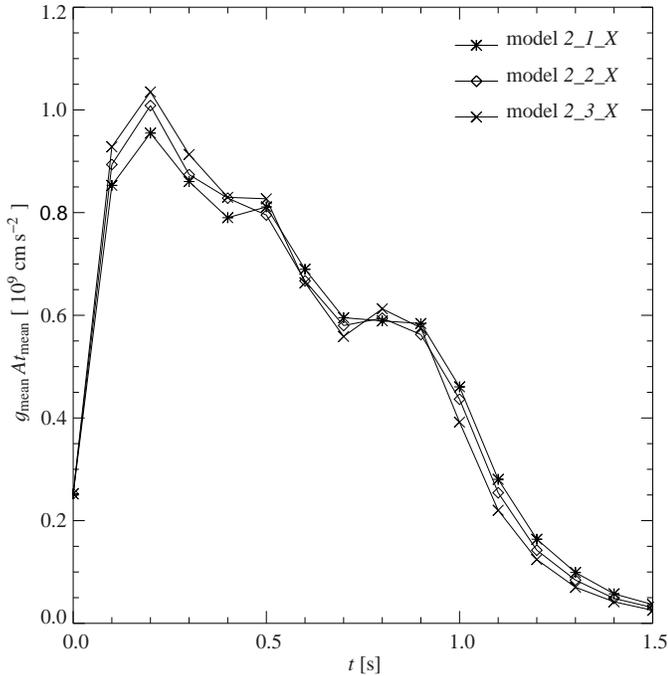}}
\caption{Mean effective gravitational accelerations experienced
  by the flame fronts for models with different C/O ratios. 
  \label{geff_co_fig}}
\end{figure}
surprisingly our models do not support the hypothesis. The energies
in our models evolve similar in the first part of the explosion and
no enhanced turbulent flame propagation is visible regardless of the
carbon mass fractions. The similar temporal evolutions of the
energetics in our models correspond to a striking similarity in the 
flame morphology.  Fig.~\ref{grid_co_fig} illustrates the flame
front in models with different C/O ratios at $t = 1.0 \,
\mathrm{s}$. The extent of the burnt volumes is comparable. The
similarities in the first part of the energy generation as well as
in the flame morphologies and flame propagations indicate that the
large scale buoyancy effects, which feed energy into the turbulent
cascade and thereby drive the flame propagation, are comparable in
models with different
C/O ratio. The buoyant velocities can be estimated from the relation
\begin{equation}
v_\mathrm{buoy} \sim \sqrt{\mathit{At}\, g\, L}
\end{equation}
for a single non-burning rising bubble of size $L$ subject to a
gravitational acceleration $g$ \citep{davies1950a}. The Atwood number $At$ characterizes the contrast
between the density inside the bubble ($\rho_\mathrm{i}$) and outside
it ($\rho_\mathrm{o}$):
\begin{equation}
\mathit{At} =
\frac{|\rho_\mathrm{o} - \rho_\mathrm{i}|}{\rho_\mathrm{o} +
  \rho_\mathrm{i}}. 
\end{equation}   
In a supernova explosion, the situation is, of 
course, much more complex since bubbles burn and will
merge. Nevertheless, it
is clear that the effective gravitational
acceleration ($\mathit{At} \, g$) determines the flame propagation
velocity -- not only directly on the largest scales but also over the
mechanism of the
interaction of the flame with generated turbulence. 
This effect can only be revealed in multidimensional calculations, as
presented here.
Fig.~\ref{geff_co_fig} shows the
mean effective gravitational acceleration ($\mathit{At} \, g$)
experienced by the flame
front. Only minor differences are visible here. 
The data point at $t = 0.0 \, \mathrm{s}$ is unphysical, since the
material behind the flame had not been burnt at this instant and thus
there is no density contrast over the flame yet. With temporal
evolution there is a competition between a rapidly decreasing
gravitational acceleration due to the expansion of the star in the
explosion and an increasing density contrast over the flame
along with lower fuel densities \citep[cf.][]{timmes1992a}. As seen from
Fig.~\ref{geff_co_fig}, finally the decreasing gravitational
acceleration dominates this competition.

Because of the very similar evolutions of the large-scale buoyancy
effects, there are little differences in  the evolutions of the 
turbulent energies in models with varying carbon mass fractions (see
Fig.~\ref{eturb_fig}\footnote{The values of $E_\mathrm{turb}$ are
not significant at late times since those are derived from the subgrid
energy which depends on the length of the grid cells. In the outer
regions which the flame enters at late times, those are elongated and
therefore $E_\mathrm{turb}$ rises again after reaching a peak at $t
\sim 0.65 \, \mathrm{s}$ and $t \sim 1.05 \, \mathrm{s}$ for $\rho_c =
2.6 \times 10^9 \,\mathrm{g}\, \mathrm{cm}^{-3}$ and $\rho_c = 1.0
\times 10^9 \,\mathrm{g}\, \mathrm{cm}^{-3}$, respectively. For a
uniform grid it would be expected to monotonically decrease after
these peaks \citep[cf.][]{roepke2004d}.}).

\begin{figure}[t]
\centerline{
\includegraphics[width = \linewidth]
  {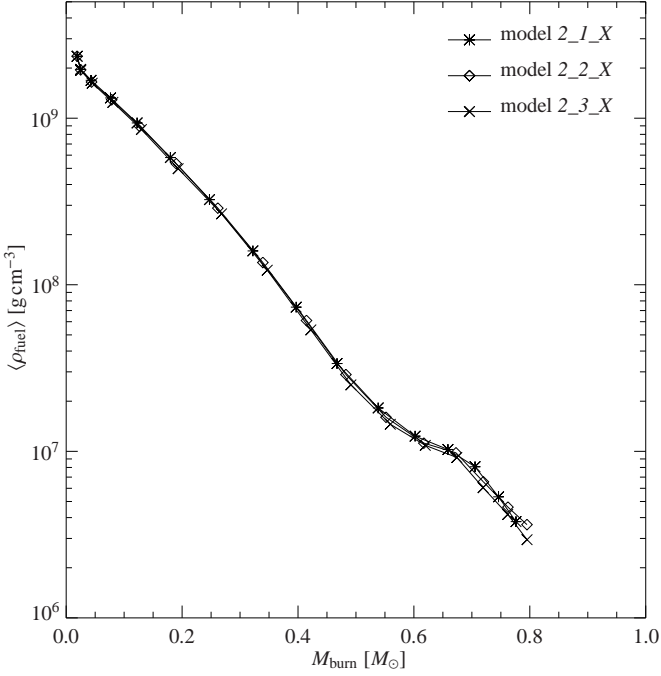}}
\caption{Fuel densities at the mean flame locations as a function of 
  the mass of ``ashes'' behind the flame front for models with
  different carbon mass fraction of the progenitor material. 
  \label{mb_co_fig}}
\end{figure}

\begin{figure}[t]
\centerline{
\includegraphics[width = \linewidth]
  {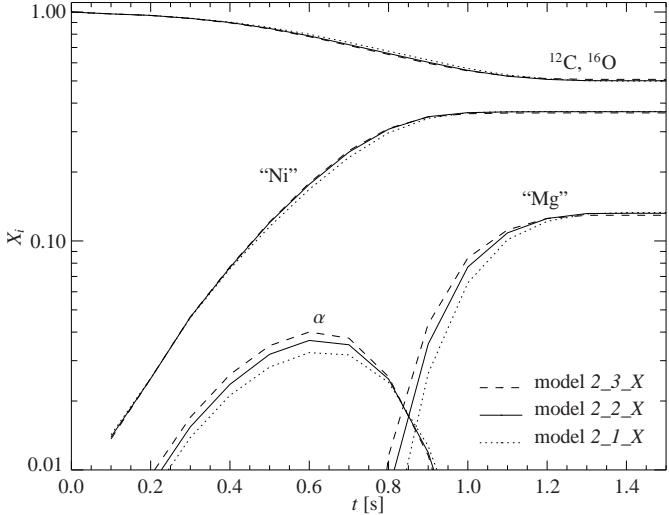}}
\caption{Temporal evolution of the chemical composition in  models with
  different carbon mass fraction of the progenitor material.
  \label{evalmass_fig}}
\end{figure}

If the larger energy from burning carbon-rich fuel was directly
converted to work expanding the ashes, buoyancy effects and an
acceleration of the turbulent flame propagation should be observable in
our simulations. The only way to bypass these effects is that the energy
is buffered in a larger fraction of $\alpha$-particles present in the
NSE. This fraction indeed increases with higher temperatures and the
consequences are twofold. First, the binding energy of the ashes is
lowered and less energy is released from thermonuclear
burning. Second, distributing the energy on an increased number of
particles in the ashes decreases their temperature. Both effects lead
to an increase in the density of the ashes which suppresses the
buoyancy effects. Hence the turbulent flame propagation
velocity in carbon-rich fuel models is lowered to values comparable to
those found in oxygen-rich fuel simulations. As a consequence similar
masses of fuel are burnt at particular fuel densities. To corroborate
this, we plot the fuel density at the average flame front location over the
burnt mass in Fig.~\ref{mb_co_fig}. 

\begin{figure}[t]
\centerline{
\includegraphics[width = \linewidth]
  {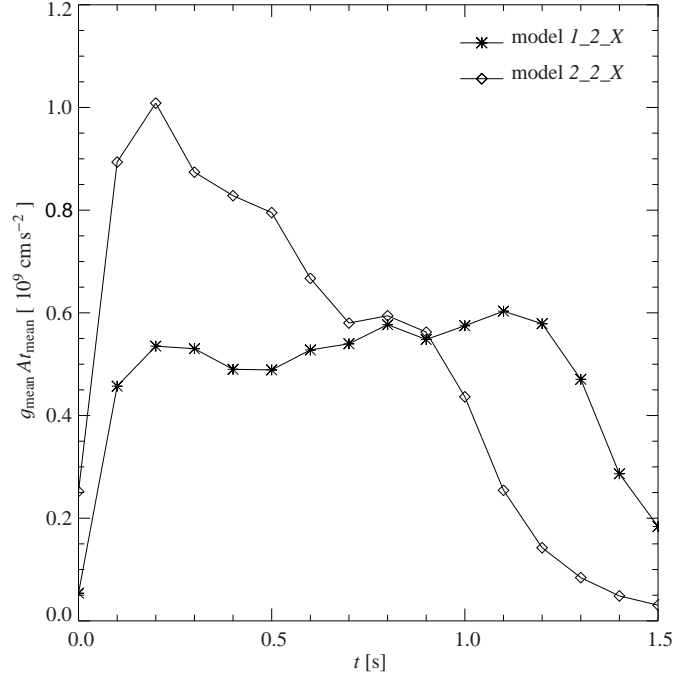}}
\caption{Mean effective gravitational accelerations experienced
  by the flame fronts for models with different central densities. 
  \label{geff_rc_fig}}
\end{figure}

Fig.~\ref{evalmass_fig} proves that the effect of energy buffering in
the $\alpha$-particles indeed applies for our
models, here shown for the models with a central density of $2.6 \,
\mathrm{g} \, \mathrm{cm}^{-3}$. Between $0.2 \, \mathrm{s}$ and $0.9\,
\mathrm{s}$ the ashes contain significant amounts of $\alpha$-particles. 
The maximum mass fraction of $\alpha$-particles increases by about
20\% when changing the carbon mass fraction from 0.30 to 0.62. This
effect is capable of compensating the differences in the fuel binding
energies according to the following estimate. At $t \sim
0.6\,\mathrm{s}$ (the energy generation rate peaks here, cf.\
Fig.~\ref{egen_fig}) the ashes in the model \emph{2\_2\_X}
contain about 21\% $\alpha$-particles and 79\% ``Ni''. If there was no
change in the amount of $\alpha$-particles along with changing the C/O
ration in the models, the nuclear energy release would increase for about
22\% changing $X(^{12}\mathrm{C})$ from 0.30 to 0.62. In contrast,
taking into account the observed 20\% increase in the
$\alpha$-particles in the ashes, the nuclear energy release difference
reduces to $\sim$5\%. 

This self-regulation mechanism has an important consequence. Since it
suppresses increased buoyancy effects which would otherwise arise with
larger carbon mass fractions in the fuel, similar amounts of fuel are
consumed by the flame at stages where burning terminates in
NSE. Therefore the amount of produced iron
group elements is nearly kept constant for different C/O ratios in the
fuel.

The $\alpha$-particles buffer the energy only temporarily. With further
expansion and cooling of the WD they are converted to iron group
elements (``Ni'') releasing the stored energy. This is the reason why
the energies in the models diverge at later times. Then, however, the
fuel density has dropped to values where burning terminates in
intermediate mass elements and hence the synthesis of iron
group elements is unaffected. Therefore the models with different
carbon-to-oxygen ratios produce similar amounts of iron group
elements. Interestingly, we find even a
slight decrease in the production of iron group elements for
an increasing carbon mass fraction of the model. The same holds for
the intermediate mass nuclei in the high central density models while the
trend is reversed in the models with lower central densities (cf.\
Table~\ref{energy_tab}).

\subsection{Variation of the central density}
\label{expl_dens}

\begin{figure*}[t]
\centerline{
\includegraphics[width = 0.667 \linewidth]
  {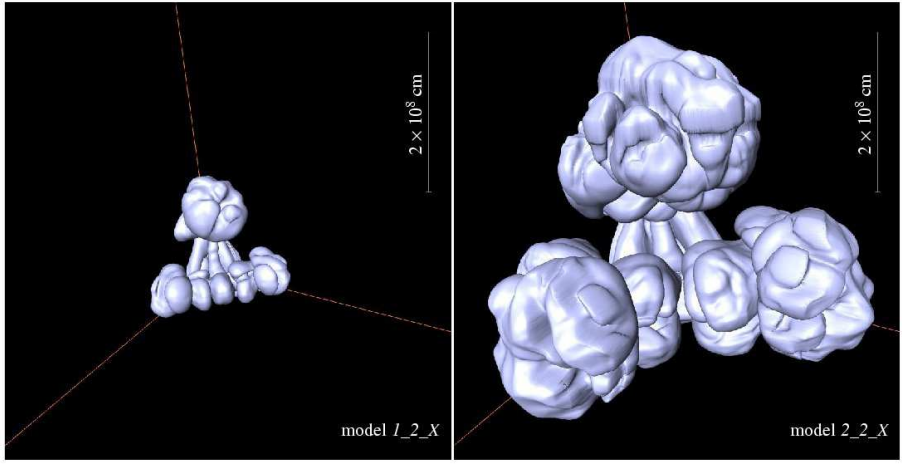}}
\caption{Flame surface of models with different central densities
  prior to ignition at $t = 1.0 \, \mathrm{s}$.
  \label{grid_rc_fig}}
\end{figure*}

For higher central densities at ignition, the explosion turns out to
be more vigorous for a fixed carbon mass fraction in the WD material
(cf.\ Fig.~\ref{etot_fig}, Table~\ref{energy_tab}). Here, the nuclear
energy releases differ about 26\% and the total energies vary about
34\%.

As for an increased carbon mass fraction, a higher density of the fuel
accelerates the laminar burning \citep{timmes1992a}. Again, this has
little impact on our models since burning proceeds laminar only in the
first few time steps. Two other effects are more significant here. 

First, for the higher central density, obviously more material is
present at sufficiently high densities so that it can potentially be
burnt to iron group elements.

Second, for higher central densities the effective gravitational
acceleration ($\mathit{At} \, g$)
experienced by the flame is higher in the first $\sim$$0.9 \,
\mathrm{s}$ (cf.\ Fig.~\ref{geff_rc_fig}). This increases the
development of flame structures resulting from the non-linear stage 
of the buoyancy instability. As a result, the turbulent cascade will
develop more quickly (cf.\ Fig.~\ref{eturb_fig}) and the
turbulence-induced boost of of the effective
flame propagation velocity
sets in earlier. This is shown in Fig.~\ref{grid_rc_fig}, where
snapshots of the flame evolutions at fixed times for models with
different central densities at ignition are given.  

Consequently, the production of iron group elements increases with
higher central densities, while the amount of intermediate mass
elements decreases. 

\section{Nucleosynthesis}
\label{nuc}

\begin{table*}
\begin{center}
\setlength{\extrarowheight}{2pt}
\begin{tabular}{lllllll}
\hline\hline
model & $\rho_c [10^9\,\textrm{g} \, \textrm{cm}^{-3}]$& $X(^{16}\textrm{O}) $ & 
$X(^{12}\textrm{C}) $ & $X(^{22}\textrm{Ne})$ &
$Z[Z_{\odot}]$  & $M(^{56}\textrm{Ni}) [M_{\odot}]$\\
\hline
\emph{1\_1\_1} & $1.0$ & $0.7$   & $0.2875$ & $0.0125$ & $0.5$ & $0.2982$\\
\emph{1\_1\_2} & $1.0$ & $0.7$   & $0.275$  & $0.025$  & $1.0$ & $0.2876$\\
\emph{1\_1\_3} & $1.0$ & $0.7$   & $0.225$  & $0.075$  & $3.0$ & $0.2450$\\
\hline
\emph{1\_2\_1} & $1.0$ & $0.54$  & $0.4475$ & $0.0125$ & $0.5$ & $0.2966$\\
\emph{1\_2\_2} & $1.0$ & $0.54$  & $0.435$  & $0.025$  & $1.0$ & $0.2860$\\
\emph{1\_2\_3} & $1.0$ & $0.54$  & $0.385$  & $0.075$  & $3.0$ & $0.2444$\\
\hline
\emph{1\_3\_1} & $1.0$ & $0.38$  & $0.6075$ & $0.0125$ & $0.5$ & $0.2907$\\
\emph{1\_3\_2} & $1.0$ & $0.38$  & $0.595$  & $0.025$  & $1.0$ & $0.2805$\\
\emph{1\_3\_3} & $1.0$ & $0.38$  & $0.545$  & $0.075$  & $3.0$ & $0.2403$\\
\hline
\emph{2\_1\_1} & $2.6$ & $0.7$   & $0.2875$ & $0.0125$ & $0.5$ & $0.3115$\\
\emph{2\_1\_2} & $2.6$ & $0.7$   & $0.275$  & $0.025$  & $1.0$ & $0.2999$\\
\emph{2\_1\_3} & $2.6$ & $0.7$   & $0.225$  & $0.075$  & $3.0$ & $0.2544$\\
\hline
\emph{2\_2\_1} & $2.6$ & $0.54$  & $0.4475$ & $0.0125$ & $0.5$ & $0.3163$\\
\emph{2\_2\_2} & $2.6$ & $0.54$  & $0.435$  & $0.025$  & $1.0$ & $0.3046$\\
\emph{2\_2\_3} & $2.6$ & $0.54$  & $0.385$  & $0.075$  & $3.0$ & $0.2592$\\
\hline
\emph{2\_3\_1} & $2.6$ & $0.38$  & $0.6075$ & $0.0125$ & $0.5$ & $0.3174$\\
\emph{2\_3\_2} & $2.6$ & $0.38$  & $0.595$  & $0.025$  & $1.0$ & $0.3065$\\
\emph{2\_3\_3} & $2.6$ & $0.38$  & $0.545$  & $0.075$  & $3.0$ & $0.2608$\\
\hline
\end{tabular}
\end{center}
\caption{$^{56}$Ni masses synthesized according to the nucleosyntehsis
  postprocessing.\label{ni_results_tab}}
\end{table*}

\begin{figure*}[ht]
\centerline{
\includegraphics[width = 0.75\linewidth]
  {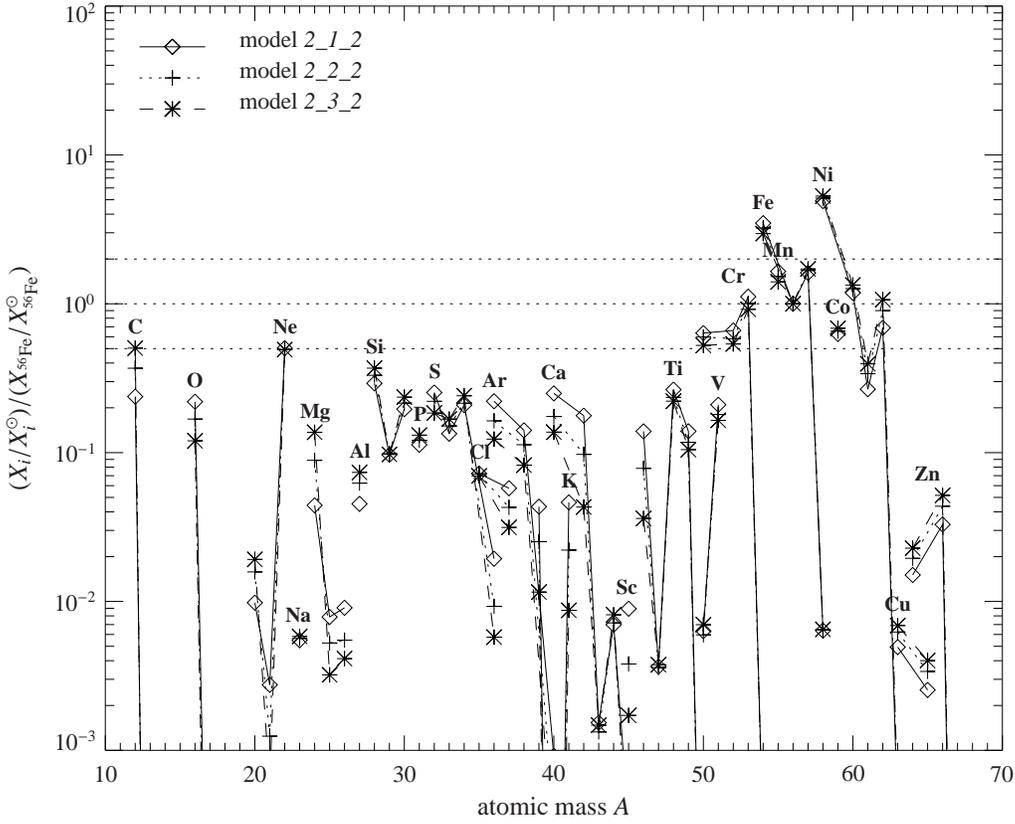}}
\caption{Final abundances for models with different C/O ratios.
  \label{abundance_co_fig}}
\end{figure*}

\begin{figure*}[ht]
\centerline{
\includegraphics[width = 0.75\linewidth]
  {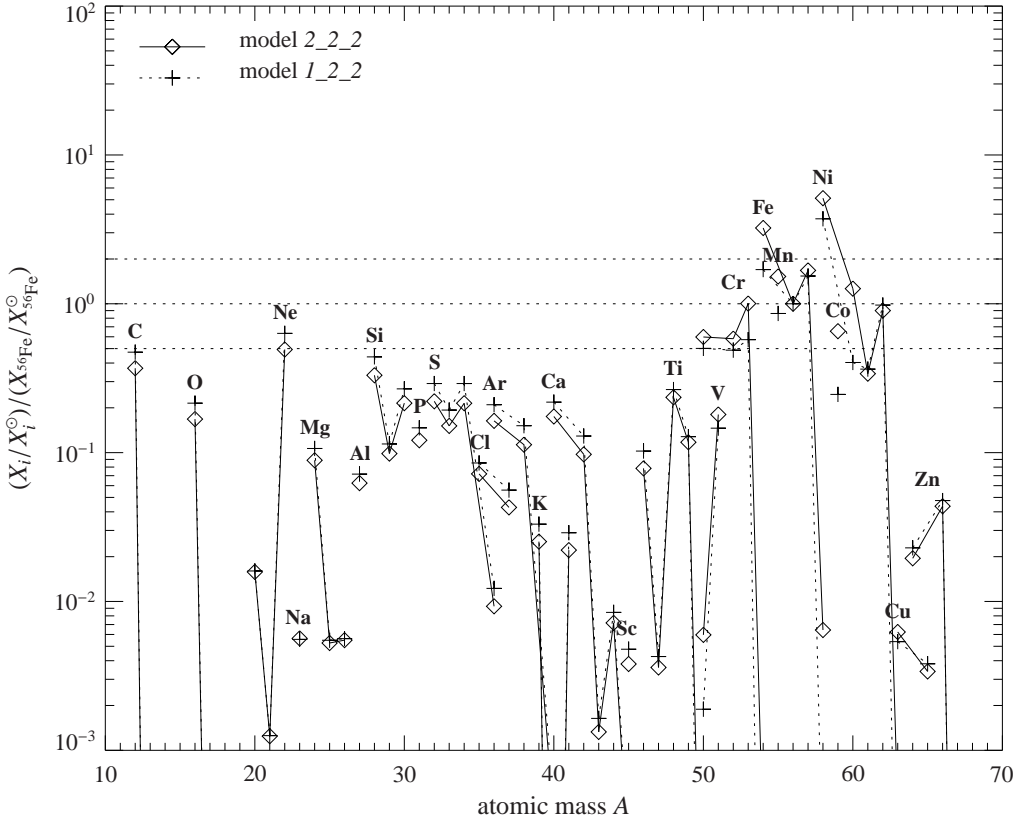}}
\caption{Final abundances for models with different central densities.
  \label{abundance_rc_fig}}
\end{figure*}

\begin{figure*}[ht]
\centerline{
\includegraphics[width = 0.75\linewidth]
  {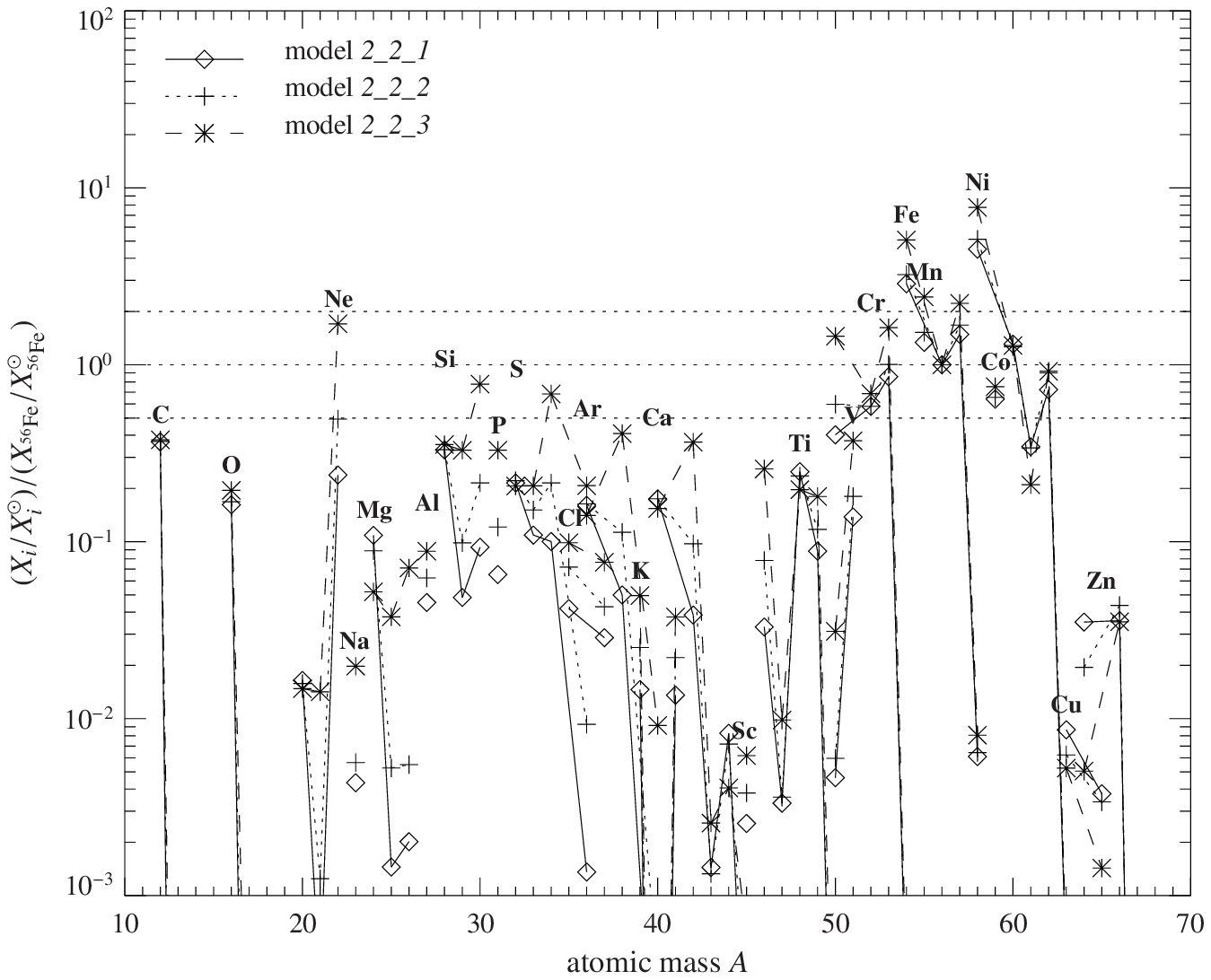}}
\caption{Final abundances for models with different metallicities.
  \label{abundance_z_fig}}
\end{figure*}

After the nucleosynthesis postprocessing the abundances of the
individual isotopes in the ashes are unveiled. Additionally, the
parameter of the progenitor's metallicity comes into play here by
assuming a certain fraction of the material to be composed of
$^{22}$Ne. 

The lightcurves of SNe Ia are powered by the radioactive decay of
$^{56}$Ni and $^{56}$Co. Therefore their peak luminosities are
determined by the nucleosynthetic yields of the explosions rather than
the energetics. Consequently, the nucleosynthetic postprocessing of
explosion models can shed some light on the observed diversity of SN
Ia events. $^{56}$Co decay is slow and thus the peak luminosity is a
function of the produced amount of $^{56}$Ni. A compilation of the
$^{56}$Ni masses derived from all
our models by nucleosynthetic postprocessing can be found in
Table~\ref{ni_results_tab}.

Although we will focus here on the effect of initial parameters on the
production of $^{56}$Ni, we will first discuss the overall picture of
the nucleosynthesis yields.

\subsection{The final yields}

The freeze-out masses after completion of the $\beta$-decays are
plotted in Figs.~\ref{abundance_co_fig} to \ref{abundance_z_fig} for
different models. Here the usual normalization to the solar abundances
and the $^{56}$Fe mass fraction was applied. Values are given in
Table~\ref{final_yields_tab} in the online material.

Fig.~\ref{abundance_co_fig} shows a comparison between models with
different carbon mass fractions for fixed $\rho_c = 2.6 \times 10^9 \,
\mathrm{g} \, \mathrm{cm}^{-3}$ and for solar metallicity. Obviously,
the carbon mass fraction has only little effect on the final
abundances. Though some variation is visible for the intermediate mass
elements (Mg to Ca), there is practically no difference in the iron
group yields for the different models. This is expected from the
analysis of the explosion process in the previous section. Due to the
energy buffering in the $\alpha$-particles, burning to NSE
consumes almost identical masses of fuel, while the recombination of
the $\alpha$-particles at the end of complete NSE burning leads to an
additional energy release that varies with the C/O ratio. Therefore
the incomplete burning in the models that follows burning to NSE
proceeds differently in the various models.

The variations in the final yields due to different central densities are illustrated in
Fig.~\ref{abundance_rc_fig}. Here, the models \emph{X\_2\_2} are
plotted, i.e. the C/O ratio is fixed to 0.81 and the metallicity is
solar. The model with lower central density produces more intermediate
mass elements, but the variations are small. In contrast, for higher central
densities, there is a visible increase in the abundances of
iron group isotopes, viz. titanium, vanadium, chromium, manganese,
iron, cobalt, and nickel. The two effects that contribute to an
increased mass consumption in complete NSE burning with higher
$\rho_c$ were discussed in Sect.~\ref{expl_dens}. The resulting final
yields are a natural consequence of these effects.

Changes in the progenitor's metallicity resulting in different
abundances of $^{22}$Ne in the WD material have a large impact on the
final yields. To illustrate this influence, we consider the models
\emph{2\_2\_X} for $0.5\,Z_\odot$, $1.0\, Z_\odot$, and $3.0\,
Z_\odot$ (cf.\ Fig.~\ref{abundance_z_fig}). The variation of the
$^{22}$Ne abundance is obvious and caused by the representation of the
progenitor's metallicity in the different mass fractions of that
isotope in our simulations. The production of chromium, manganese and
iron isotopes is increased for higher metallicity, especially for
isotopes with two more neutrons than the symmetric nuclei ($^{54}$Fe,
$^{58}$Ni). An exception is $^{56}$Fe which was used to normalize the
abundances. This trend holds analogously for intermediate mass
elements. 
In particular, one observes a higher abundance of $^{26}$Mg, $^{30}$Si,
$^{34}$S, $^{38}$Ar and $^{42}$Ca with increased metallicity. 
Comparing the models \emph{2\_2\_2} and \emph{2\_2\_3} the change is
by a factor of 11 for $^{26}$Mg and a factor of approximately $3$ for
the other isotopes (cf.\ Table~\ref{final_yields_tab} in the online
material). The other
models not present in the table give similar factors for identical
metallicities. The increase of neutron-rich isotopes is caused by the
fact that a higher progenitor's metallicity results in an increased
$^{22}$Ne mass fraction, which serves as a source of a neutron excess.

\subsection{Impact of the C/O ratio on the $^\textsf{56}$Ni mass}

Contrary to the previous section, we analyze here the 
nucleosynthesis yields right after the explosion. The production of
radioactive species is given in Table~\ref{radio_tab} in the online
material.

\begin{figure}[t]
\centerline{
\includegraphics[width = \linewidth]
  {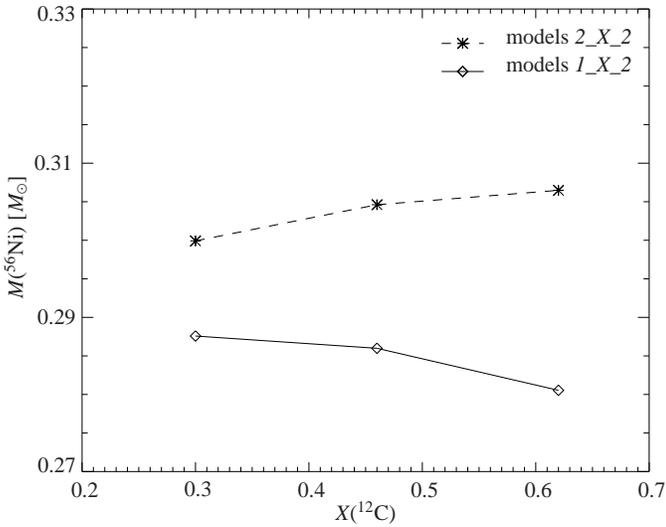}}
\caption{$^{56}$Ni production depending on the carbon mass fraction
  for models with different central densities and solar
  metallicity. \label{nickel_co_fig}}
\end{figure}

Fig.~\ref{nickel_co_fig} shows the $^{56}$Ni
production for the models in dependence of the carbon mass fraction of
the progenitor. The central densities are fixed to $\rho_c = 1.0
\times 10^9 \,
\mathrm{g} \, \mathrm{cm}^{-3}$ and $\rho_c = 2.6 \times 10^9 \,
\mathrm{g} \, \mathrm{cm}^{-3}$, respectively. The metallicities of
the models shown 
here are set to $Z=Z_\odot$. We note only minor changes in the
$^{56}$Ni masses (about 2\%) for both central densities, which is
not surprising given the small variations in the flame morphology and
advancement discussed in Sect.~\ref{expl_dens}. 

It should be noted in Fig.~\ref{nickel_co_fig} that the trend of
$^{56}$Ni production has opposite directions for different central
densities. While this feature is in accordance with the total production
of iron group elements in the explosion models for low central
densities, it is reversed for the high central density case (cf.\
Table~\ref{energy_tab}). In order to check whether
an under-representation of NSE-material in tracer particles in the low
density case was the origin, we recalculated these models with the
number of tracers increased to $35^3$. The trend of decreasing
$^{56}$Ni production with higher carbon mass fraction was weaker, but
had still the same direction. Since the variations are at the percent
level, it is beyond the accuracy of our models to judge whether the
trend is of physical nature or an artifact of our simulation.   

The result of the $^{56}$Ni production in the explosion phase being
largely independent of the carbon mass fraction supports the
conjecture of \citet{roepke2004c}
that the peak luminosity of SNe Ia will be only marginally affected by the
carbon-to-oxygen ratio of the progenition WD star. This conjecture was
only based on the cumulative production of all iron group elements and
is now confirmed by the derivation of the exact amounts of $^{56}$Ni
via nucleosynthetic postprocessing.

\subsection{Impact of the central density on the $^\textsf{56}$Ni mass}

For fixed C/O ratios of 0.81 and solar metallicities our models
produce $0.286\,M_\odot$ of $^{56}$Ni for $\rho_c = 1.0\times
10^9\,\mathrm{g}\,\mathrm{cm}^{-3}$ and $0.305\,M_\odot$ of $^{56}$Ni
for $\rho_c = 2.6\times 10^9\,\mathrm{g}\,\mathrm{cm}^{-3}$, i.e.\
from the lower to the higher central density the $^{56}$Ni production
increases for 7\%. These changes go along with the higher overall
production of iron group nuclei at higher central densities (cf.\
Table~\ref{energy_tab}). The
reasons for this effect have been discussed in Sect.~\ref{expl_dens}.

Although somewhat larger than the changes found in the case of varying
carbon mass fraction, the effect is still rather small. However,
our study covers only parts of the
effects resulting from changing the central densities of the
models. With a further increasing central density, electron captures
will become important and the $^{56}$Ni production is expected to
decrease while the total mass of iron group elements should still
increase. Unfortunately, in the current study this effect could not
be consistently modeled, but it will be addressed in forthcoming work.

\subsection{Impact of the metallicity on the $^\textsf{56}$Ni mass}

\begin{figure}[t]
\centerline{
\includegraphics[width = \linewidth]
  {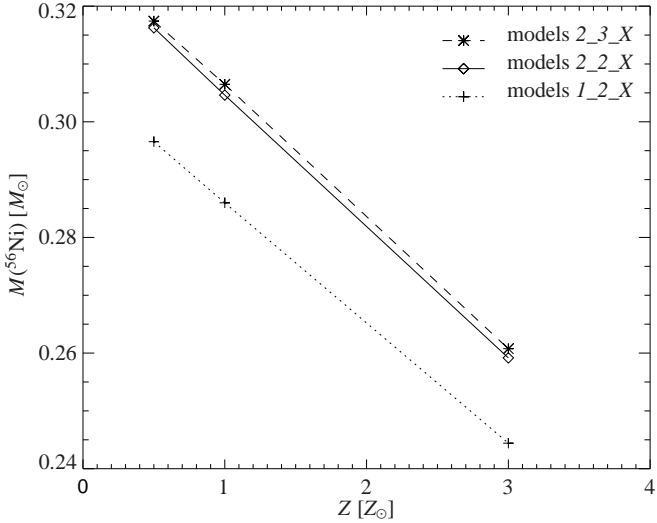}}
\caption{$^{56}$Ni production depending on the progenitor's
  metallicity for models with different central densities C/O
  ratios. \label{nickel_z_fig}} 
\end{figure}

\citet{timmes2003a} proposed an analytic model for the effect of the
progenitor's metallicity on the $^{56}$Ni production in SN Ia
explosions. Their reasoning is based on  the assumptions that most of 
the $^{56}$Ni is produced between the $0.2 M_\odot$ and the $0.8
M_\odot$ mass shells in NSE and that the $Y_e$ is constant during
burning in that region. This is motivated by one-dimensional
models. Furthermore, they take into account only the species with the
highest mass fraction; in a first step $^{56}$Ni and $^{58}$Ni. Under
these assumptions they derive a linear correlation between Z and the 
produced $M(^{56}\mathrm{Ni})$:

\begin{equation}
\frac{M_{^{56}\mathrm{Ni}}(\tilde{Z})}{M_{^{56}\mathrm{Ni}}(\tilde{Z} = 0)} = 1 - 0.057 \tilde{Z},
\label{linear1}
\end{equation}
with $\tilde{Z} = Z/Z_\odot$.
This equation is obtained from combining the equations
\begin{equation}
\frac{M_{^{56}\mathrm{Ni}}}{M_{^{56}\mathrm{Ni}}(\tilde{Z} = 0)} = 58 Y_e -28,
\label{lin2}
\end{equation}
resulting from conservation of mass and charge,
\begin{equation}
X(^{22}\mathrm{Ne}) = 22 \left(\frac{X(^{12}\mathrm{C})}{12} +
                               \frac{X(^{14}\mathrm{N})}{14} + 
                               \frac{X(^{16}\mathrm{O})}{16} \right),
\label{ignore}
\end{equation}
approximating the $^{22}$Ne abundance resulting from the metallicity
of the ZAMS progenitor, and
\begin{equation}\label{ye-init}
\begin{split}
Y_e = \frac{10}{22} & X(^{22}\mathrm{Ne}) + \frac{26}{56}
X(^{56}\mathrm{Fe}) \\
+ \frac{1}{2} & \left [ 1 - X(^{22}\mathrm{Ne}) - X(^{56}\mathrm{Fe})
\right ],
\end{split}
\end{equation}
giving the initial $Y_e$ of the white dwarf right before the explosion
under the assumption of a uniform distribution of $^{22}$Ne and
$^{56}$Fe. When the presence of $^{54}$Fe is
taken into account the factor 0.057 in Eq.~(\ref{linear1}) changes to
0.054 \citep[cf.][]{timmes2003a}.

For a comparison of this analytic prediction with our models we set
$X(^{56}\mathrm{Fe}) = 0$ in Eq.~(\ref{ye-init}) since here the
initial $Y_e$ is determined by $^{22}\mathrm{Ne}$ only. 
We set $X(^{22}\mathrm{Ne}) = 0.025 \tilde{Z}$, where
$\tilde{Z} = Z/ Z_\odot$. If Eq.~(\ref{ye-init}) is now substituted into
Eq.~(\ref{lin2}), the following equation is obtained
\begin{equation}
\frac{M_{^{56}\mathrm{Ni}}(\tilde{Z})}{M_{^{56}\mathrm{Ni}}(\tilde{Z} = 0)} = 1 - \frac{58}{22}\,0.025 \tilde{Z}.
\label{our-lin}
\end{equation}

To compare this linear dependence of the $^{56}$Ni mass 
on $\tilde{Z}$ with our simulations, a linear regression
following
\begin{equation}
M_{^{56}\mathrm{Ni}}(\tilde{Z}) = M_0 + m \tilde{Z}
\label{ansatz}
\end{equation} 
was applied to our data. Here $M_0$ denotes the extrapolated value
$M(^{56}\mathrm{Ni})$ at $\tilde{Z} = 0$. Values for $m$ and $M_0$ for
the different models are given in Table~\ref{metal_tab}. The coefficient of
correlation is 0.9999 in all cases, which suggests a good agreement of
our data with a linear dependence, but, of course, more data points
would be desirable for a definite statement.

\begin{table*}
\centering
\caption{Fit parameters according to Eq.\ref{ansatz} from our models.
\label{metal_tab}}
\setlength{\extrarowheight}{2pt}
\begin{tabular}{rllllll}
\hline\hline
model & \emph{1\_1\_X} & \emph{1\_2\_X} & \emph{1\_3\_X} &
\emph{2\_1\_X} & \emph{2\_2\_X} & \emph{2\_3\_X} \\
\hline
$-m$ &  0.0213  & 0.0209  & 0.0201  & 0.0228
& 0.0228 & 0.0227 \\
$M_0$ $[M_{\odot}]$ & 0.3089  & 0.3070  & 0.3007  & 0.3228
& 0.3275 & 0.3290 \\
$-m/M_0$ $[M_\odot^{-1}]$ & 0.0690 & 0.0681 & 0.0668 & 0.0706 &
0.0696 & 0.0690 \\
\hline
\end{tabular}
\end{table*}

To compare the slope of (\ref{our-lin}), i.e. $-0.0659$, with the fits
to our data, we give values for $-m/M_0$ in Table~\ref{metal_tab}. The
agreement is reasonable keeping in mind that Eqs.~(\ref{linear1}) and
(\ref{our-lin}) were derived by assuming 
$^{56}$Ni and $^{58}$Ni to be the two most abundant isotopes in NSE
with a constant $Y_e$. However, a significant amount of  $^{56}$Ni
seems to be produced in regions where
the assumption of constant $Y_e$ breaks down.

Thus, the analytical model introduced in \citet{timmes2003a} 
provides an excellent explanation for the effect of the
metallicity. Based on models different
from the ones applied here this was recently confirmed by
\citet{travaglio2005a}.

\section{Conclusions}
\label{concl}

In the present paper the impact of several progenitor parameters on
three-dimensional SN Ia explosion models has been studied for the
first time in a systematic way. Here, we investigated the effects of
the progenitor's central density, its carbon mass fraction, and its
metallicity. Of course, there may be several other parameters that
possibly affect the light curve of SNe Ia (rotation of the
progenitor, morphology of the ignition spot(s), a delayed detonation
at varying densities, asphericities etc.), which were not addressed
in the present survey. 

A first important point to note is that our numerical implementation
as well as the underlying astrophysical model are evidently robust
against variations of the initial conditions to a reasonable
degree. On the one hand, the variations in the resulting features are
relatively small. A deviation in orders of magnitude should have been
reason for concern, but all our models seem to be well-behaved. On the
other hand, the model is not too robust in the sense that
variations of the initial parameters do show effects on the
results, i.e.\ an intrinsic variability is preserved. The degrees
of freedom expected for a SN Ia explosion 
are at least not entirely artificially suppressed in our model.
Hence our model fulfills the requirements 2, 3, and 4 stated by
\citet{hillebrandt2000a}. 

Another point is the absolute scale of the results.
Given the limited resources of
computational time and storage space, we had to restrict the models to
a resolution of $[256]^3$ grid cells per octant. Although such models
reach numerical convergence in global characteristics
\citep{reinecke2002c}, it is not possible to apply multi-spot 
ignition scenarios  at this resolution which would produce more
vigorous explosions. As a consequence, the explosion energy of all our
models is rather low and the $^{56}$Ni production falls short the
nickel mass of a prototype SN Ia (\citet{contardo2000a} find $0.41
M_\odot$ of $^{56}$Ni for SN1994D).  These
restrictions exclude the possibility of finding the absolute scale of
effects 
and hence requirement 1 of
\citet{hillebrandt2000a} is not met in the current study. However,
there is a fair chance that models with more elaborate initial flame
representations will agree better with the absolute values of observed
quantities \citep[see e.g.][]{travaglio2004a}.
Nevertheless the present parameter study should reveal the correct
trends of the
variation of SN Ia properties.

A major uncertainty lies in the range of variation of the
progenitor parameters. Although we applied values that are common in
literature, our parameter space is not derived from a realistic stellar
evolution of the progenitor. 

Keeping this in mind, the maximum variation in $^{56}$Ni of about 27\%
found in our parameter study can be regarded as a strong hint that the
variations of the progenitor properties taken into account here
provide a significant contribution to the scatter in SN Ia
luminosities. However, it seems unlikely that these are sufficient to
explain the full range of diversities in ``Branch normal'' SNe Ia. Of course, more
elaborate models are required to assess this.

Regarding the diversity of $^{56}$Ni production in our models
resulting from the variation of the initial parameters, the following
trends were found:
\begin{itemize}
\item The \emph{progenitor's carbon-to-oxygen ratio} has only little impact
  on the amount of produced $^{56}$Ni. This is in strong contrast to
  the common assumption that the C/O ratio be a major source of
  luminosity variation in SN Ia explosions. The ``working hypothesis''
  of \citet{umeda1999b} could not be confirmed by our models. The
  reason for this effect could be unveiled by our three-dimensional
  simulations. Since flame propagation in the deflagration stage is
  mainly determined by the turbulent motions of the material, the
  explosion dynamics is not altered as long as the buoyancy effects
  that generate the turbulence are comparable. This is given in our
  models at stages of iron group nuclei synthesis.  Different
  energy releases resulting from differences in the fuel binding energies
  are compensated by a varying amount of $\alpha$-particles present in
  the ashes. These buffer the temperature of the ashes and thus the
  densities are not altered substantially ensuring the same
  buoyancies. Consequently, the explosion dynamics is similar
  in the stages of iron group element synthesis for models with
  different C/O ratio in the fuel resulting in a small variation of
  the produced $^{56}$Ni of only about 2\%.
\item The \emph{central density} affects the $^{56}$Ni production. The
  variation found in our models amounts to about 7\%. This is
  explained by the fact that for higher central densities more
  material is burned under conditions where iron group elements are
  produced. Moreover, a higher central density increases the mean
  gravitational acceleration experienced by the flame front and thus
  enhances the generation of turbulence thereby accelerating the flame
  propagation. Due to this effect even more material is processed at
  higher densities where the reactions terminate in iron group
  elements.
\item A greater effect (assuming that our parameter space is reasonable)
  was found for a variation of the \emph{metallicity} in the nuclear
  postprocessing. By varying the $^{22}$Ne mass fraction from 0.5 to 3
  times solar, a variation of the produced $^{56}$Ni mass of about
  20\% was found. Our models were consistent with the analytical
  prediction of a linear relation between  the metallicity and
  $X(^{56}\mathrm{Ni})$ by \citet{timmes2003a}.
\end{itemize}

The effects of varying C/O ratios and central densities of the
progenitor on the supernova explosion are based on effects of the
turbulent flame propagation and can thus only be revealed by
three-dimensional models.

However, we have to emphasize an important limitation of the
results. Our analysis addresses
only changes in the explosion process itself. For comparability of the
simulations we assumed identical initial flame configurations. The
ignition process, however, may be influenced by the carbon-to-oxygen
ratio of the progenitor \citep{woosley2004a}. Since different initial
flames can have large impact on the explosion dynamics
\citep[e.g.][]{reinecke2002d, gamezo2003a, calder2004a, roepke2005b}, the
C/O ratio may still be an important parameter via this mechanism.

We emphasize an incompleteness of the present survey towards higher
densities, at which
electron captures in the ashes become important. These shift the
burning products to neutron-rich isotopes, favoring $^{58}$Ni instead
of $^{56}$Ni. This effect would be taken into account in our
postprocessing procedure, however electron captures may also become
dynamically important with increasing central densities, since they
reduce the electron pressure in the ashes.
 Unfortunately, this effect could not
consistently be modeled in the current study. The explosion model
assumes $Y_e = 0.5$. Effects of higher central
densities will be addressed in forthcoming investigations.

\subsection{Comparison with one-dimensional models}

The effect of a variation in the carbon mass fraction of the
progenitor on the produced $^{56}$Ni mass was studied by
\citet{hoeflich1998a}. They applied a one-dimensional delayed
detonation model. For a central density of $2.6 \times 10^9 \,
\mathrm{g} \, \mathrm{cm}^{-3}$, solar metallicity and a presumed
deflagration-to-detonation transition at a density of $2.7 \times 10^7 \,
\mathrm{g} \, \mathrm{cm}^{-3}$ they calculated a model with a C/O
ratio of 1/1 (DD21c in their notation) and a model with C/O reduced to
2/3. Here they find a decrease of the produced $M(^{56}\mathrm{Ni})$
of about 14\%. Assuming that a transition to detonation at such
low densities as applied here does not alter the production of iron
peak elements, a comparison with our models is possible. However, the
results of \citet{hoeflich1998a} are in contrast with ours. This
may be mainly due to the fact that the modeling of the correct
implications of the C/O ratio on the explosion results requires
an accurate description of the multidimensional effects that dominate
the flame propagation.

\citet{bravo1993a} investigated the impact of the ignition density on
the $^{56}$Ni production for one-dimensional deflagration models. For
models with a central density of $2.5 \times 10^9 \, \mathrm{g} \,
\mathrm{cm}^{-3}$ (R2 in their notation) and $4.0 \times 10^9 \, \mathrm{g} \,
\mathrm{cm}^{-3}$ (R4) they find differences of about 7\% which is in
good agreement with our results.

Although our results regarding the change in $^{56}$Ni production
varying the metallicity are in good agreement with the analytical
prediction by \citet{timmes2003a} and with the study by
\citet{travaglio2005a}, they disagree with the findings of
\citet{hoeflich1998a}. They report an only
$\sim$5\% effect on the $^{56}$Ni production changing the metallicity
from $0.1 Z_\odot$ to $10 Z_\odot$. Contrary to this, the result of
\citet{iwamoto1999a} that an increase of the metallicity from zero to
solar decreases the $^{56}$Ni production for about 8\% is consistent
with our results. 

\subsection{Cosmological significance}

In order to discuss the
cosmological significance of our results, we take
on a very simplistic view on the mechanism of the light
curve. Following ``Arnett's rule'' \citep{arnett1982a} we assume that
the total mass of
$^{56}$Ni immediately determines the peak luminosity of the SN Ia
event. Furthermore we assume that a larger energy released in the
explosion leads to a more rapid decline of the light curve
\citep{pinto2000a}. It has to be noted, however, that this may
only be a second-order effect in the light-curve shape. The main
parameter here is the opacity given by the distribution of heavy
elements \citep{mazzali2001a}. This can only be adequately addressed in
detailed synthetic light curve calculations and will be ignored here.

In the context of this simplification, we note that none of the tested
parameters reproduces the peak
luminosity-light curve shape relation by lowering the produced
$^{56}$Ni mass accompanied by an increased energy release. While the
carbon-to-oxygen ratio of the progenitor has little effect on the peak
luminosity, it could alter the width of the light curve. The opposite
holds for the progenitor's
metallicity. Here, the peak luminosity can be vary by about 20\%,
but the explosion dynamics is unaffected. The central density
prior to the ignition changes both the $^{56}$Ni production and the
energy release. Unfortunately our study is incomplete here. At higher
values for the central density, the produced $^{56}$Ni mass could
decrease due to electron captures while the energy release may still
increase. This has to be tested in forthcoming studies.

Another aspect is that we have ignored the interrelation of the
parameters by stellar evolution here. This, however, predicts a lower
C/O ratio for higher metallicities \citep[cf.][]{umeda1999a}. The
effects of both parameters in this combination may possibly reproduce
the trend of the peak luminosity-light curve shape relation.

The final conclusions on the cosmological significance of the
variations in the explosions found in the present study need to be
drawn on the basis of synthetic light curves derived from our
models. This is subject of a subsequent publication. 

\begin{acknowledgements} 
This work was supported in part by the European Research
Training Network ``The Physics of Type Ia Supernova Explosions'' under
contract HPRN-CT-2002-00303.
\end{acknowledgements}

\Online

\begin{longtable}{c c c c c c c }
\caption{\label{final_yields_tab} Synthesized mass $(M_{\odot})$ for different models as
  plotted in Figs.~\ref{abundance_co_fig}, \ref{abundance_rc_fig}, and
  \ref{abundance_z_fig}.}\\
\hline\hline
 Species & \multicolumn{6}{c}{model} \\
& \emph{2\_1\_2} & \emph{2\_2\_2} & \emph{2\_3\_2} & \emph{1\_2\_2} & \emph{2\_2\_1} & \emph{2\_2\_3}\\
\hline
\endfirsthead
\caption{continued.}\\
\hline\hline
Species & \multicolumn{6}{c}{model} \\
& \emph{2\_1\_2} & \emph{2\_2\_2} & \emph{2\_3\_2} & \emph{1\_2\_2} & \emph{2\_2\_1} & \emph{2\_2\_3}\\
\hline
\endhead
\hline
\endfoot
$^{12}$C &1.9875E-01 &3.1231E-01 &4.2854E-01 &3.5180E-01 &3.2127E-01 & 2.7646E-01 \\
$^{13}$C &4.0934E-10 &1.5019E-10 &5.2597E-11 &1.0447E-10 &4.0874E-11 & 1.5923E-09 \\
$^{15}$N &6.6972E-11 &6.3314E-11 &1.2747E-10 &5.3309E-11 &2.3057E-10 & 1.1065E-09 \\
$^{16}$O &5.8071E-01 &4.5006E-01 &3.2308E-01 &5.0524E-01 &4.4801E-01 & 4.5299E-01 \\
$^{17}$O &3.8492E-09 &1.6084E-09 &7.9966E-10 &1.3737E-09 &4.2219E-10 & 1.6026E-08 \\
$^{18}$O &1.6529E-10 &1.8213E-10 &1.3201E-10 &1.6677E-10 &1.2570E-10 & 4.3167E-10 \\
$^{19}$F &1.1308E-10 &4.8632E-11 &1.9358E-11 &4.1208E-11 &9.5085E-12 & 4.3938E-10 \\
$^{20}$Ne&4.3781E-03 &7.1415E-03 &8.7143E-03 &6.3617E-03 &7.6873E-03 & 5.7854E-03 \\
$^{21}$Ne&3.1305E-06 &1.4348E-06 &6.6327E-07 &1.2694E-06 &3.6188E-07 & 1.4177E-05 \\
$^{22}$Ne&1.8029E-02 &1.7916E-02 &1.7980E-02 &2.0192E-02 &8.9582E-03 & 5.3749E-02 \\
$^{23}$Na&5.0496E-05 &5.2638E-05 &5.4596E-05 &4.5513E-05 &4.1890E-05 & 1.5990E-04 \\
$^{24}$Mg&6.2728E-03 &1.2767E-02 &1.9800E-02 &1.3463E-02 &1.6159E-02 & 6.4936E-03 \\
$^{25}$Mg&1.4661E-04 &9.9366E-05 &6.0991E-05 &9.1044E-05 &2.8369E-05 & 6.1613E-04 \\
$^{26}$Mg&1.9381E-04 &1.1919E-04 &8.9878E-05 &1.0706E-04 &4.5240E-05 & 1.3344E-03 \\
$^{27}$Al&7.2327E-04 &1.0101E-03 &1.1982E-03 &1.0191E-03 &7.5928E-04 & 1.2407E-03 \\
$^{28}$Si&5.2546E-02 &6.0341E-02 &6.7721E-02 &7.0398E-02 &6.2266E-02 & 5.5974E-02 \\
$^{29}$Si&9.1322E-04 &9.4254E-04 &9.3651E-04 &9.6043E-04 &4.7884E-04 & 2.7276E-03 \\
$^{30}$Si&1.2685E-03 &1.4113E-03 &1.5581E-03 &1.5451E-03 &6.3194E-04 & 4.4249E-03 \\
$^{31}$P &2.5345E-04 &2.7570E-04 &3.0006E-04 &2.9326E-04 &1.5368E-04 & 6.4987E-04 \\
$^{32}$S &2.7642E-02 &2.4441E-02 &2.0486E-02 &2.8237E-02 &2.4557E-02 & 1.9786E-02 \\
$^{33}$S &1.1883E-04 &1.3593E-04 &1.5201E-04 &1.5252E-04 &1.0148E-04 & 1.6145E-04 \\
$^{34}$S &1.0709E-03 &1.1193E-03 &1.2621E-03 &1.3314E-03 &5.4058E-04 & 3.0902E-03 \\
$^{36}$S &5.0047E-07 &2.4312E-07 &1.5159E-07 &2.8158E-07 &3.6821E-08 & 4.7111E-06 \\
$^{35}$Cl&5.0421E-05 &5.0890E-05 &4.9569E-05 &5.2873E-05 &3.0618E-05 & 6.0507E-05 \\
$^{37}$Cl&1.3557E-05 &1.0231E-05 &7.5430E-06 &1.1730E-05 &7.0770E-06 & 1.5845E-05 \\
$^{36}$Ar&4.7197E-03 &3.5313E-03 &2.6786E-03 &3.9809E-03 &3.5923E-03 & 2.6380E-03 \\
$^{38}$Ar&5.9756E-04 &4.8536E-04 &3.5572E-04 &5.7181E-04 &2.2226E-04 & 1.5216E-03 \\
$^{40}$Ar&6.3966E-09 &2.3371E-09 &1.2911E-09 &2.6198E-09 &2.6172E-10 & 5.5970E-08 \\
$^{39}$K &4.1443E-05 &2.4448E-05 &1.1254E-05 &2.8140E-05 &1.4635E-05 & 4.1644E-05 \\
$^{40}$K &1.5893E-08 &1.1857E-08 &8.8932E-09 &1.2172E-08 &3.5607E-09 & 1.5243E-08 \\
$^{41}$K &3.3541E-06 &1.6253E-06 &6.4371E-07 &1.8679E-06 &1.0319E-06 & 2.3920E-06 \\
$^{40}$Ca&4.1245E-03 &2.9254E-03 &2.3081E-03 &3.2055E-03 &3.0169E-03 & 2.2343E-03 \\
$^{42}$Ca&2.0467E-05 &1.1405E-05 &5.0761E-06 &1.3292E-05 &4.6711E-06 & 3.7149E-05 \\
$^{43}$Ca&3.7921E-08 &3.3294E-08 &3.7149E-08 &3.6033E-08 &3.7361E-08 & 5.5808E-08 \\
$^{44}$Ca&2.7296E-06 &2.8433E-06 &3.2414E-06 &2.9370E-06 &3.3912E-06 & 1.3942E-06 \\
$^{46}$Ca&1.1429E-11 &3.9578E-12 &1.7961E-12 &4.6778E-12 &3.6129E-13 & 6.5306E-11 \\
$^{48}$Ca&5.1102E-16 &4.0742E-17 &8.4593E-18 &4.7231E-17 &7.0210E-19 & 3.8800E-14 \\
$^{45}$Sc&9.5903E-08 &4.1321E-08 &1.8759E-08 &4.5593E-08 &2.8767E-08 & 5.8257E-08 \\
$^{46}$Ti&8.5527E-06 &4.8903E-06 &2.2655E-06 &5.6091E-06 &2.1275E-06 & 1.3952E-05 \\
$^{47}$Ti&2.0932E-07 &2.0933E-07 &2.1956E-07 &2.1773E-07 &2.0015E-07 & 4.9531E-07 \\
$^{48}$Ti&1.5686E-04 &1.4112E-04 &1.3379E-04 &1.3985E-04 &1.5444E-04 & 1.0256E-04 \\
$^{49}$Ti&6.2734E-06 &5.3616E-06 &4.8099E-06 &5.1362E-06 &4.1781E-06 & 7.1369E-06 \\
$^{50}$Ti&4.0498E-10 &3.9554E-10 &4.0308E-10 &3.3934E-11 &3.5590E-10 & 1.1353E-09 \\
$^{50}$V &1.6127E-09 &1.5420E-09 &1.8142E-09 &4.2838E-10 &1.2401E-09 & 6.9824E-09 \\
$^{51}$V &2.1755E-05 &1.8974E-05 &1.7403E-05 &1.3477E-05 &1.5028E-05 & 3.3895E-05 \\
$^{50}$Cr&1.2996E-04 &1.2398E-04 &1.0947E-04 &9.1295E-05 &8.6254E-05 & 2.6088E-04 \\
$^{52}$Cr&2.7097E-03 &2.4212E-03 &2.2432E-03 &1.7785E-03 &2.5029E-03 & 2.4744E-03 \\
$^{53}$Cr&5.2887E-04 &4.8179E-04 &4.4336E-04 &2.4196E-04 &4.2566E-04 & 6.7475E-04 \\
$^{54}$Cr&2.8910E-08 &2.9013E-08 &2.8687E-08 &5.3360E-11 &2.7851E-08 & 3.6635E-08 \\
$^{55}$Mn&6.0348E-03 &5.6617E-03 &5.2371E-03 &2.8042E-03 &5.1364E-03 & 7.7788E-03 \\
$^{54}$Fe&6.8344E-02 &6.4350E-02 &5.9272E-02 &2.9665E-02 &5.9137E-02 & 8.7595E-02 \\
$^{56}$Fe&3.2193E-01 &3.2643E-01 &3.2811E-01 &2.8657E-01 &3.3773E-01 & 2.8308E-01 \\
$^{57}$Fe&1.2662E-02 &1.3360E-02 &1.3723E-02 &1.0757E-02 &1.2245E-02 & 1.5385E-02 \\
$^{58}$Fe&6.4899E-06 &6.6366E-06 &6.7433E-06 &1.2656E-07 &6.5205E-06 & 7.2146E-06 \\
$^{59}$Co&5.8039E-04 &6.1402E-04 &6.4570E-04 &2.0260E-04 &6.2006E-04 & 6.1134E-04 \\
$^{58}$Ni&6.6717E-02 &7.0699E-02 &7.3404E-02 &4.5109E-02 &6.4330E-02 & 9.2729E-02 \\
$^{60}$Ni&6.3865E-03 &6.9060E-03 &7.3461E-03 &1.9357E-03 &7.3971E-03 & 6.0938E-03 \\
$^{61}$Ni&6.3095E-05 &8.1359E-05 &9.5324E-05 &7.6708E-05 &8.5555E-05 & 4.3643E-05 \\
$^{62}$Ni&5.2705E-04 &6.9753E-04 &8.2941E-04 &6.6740E-04 &5.8056E-04 & 6.1748E-04 \\
$^{64}$Ni&3.4339E-11 &3.8872E-11 &4.2017E-11 &5.6274E-14 &3.7882E-11 & 4.3591E-11 \\
$^{63}$Cu&7.8178E-07 &1.0014E-06 &1.1137E-06 &7.5785E-07 &1.4388E-06 & 7.3209E-07 \\
$^{65}$Cu&1.8541E-07 &2.5041E-07 &2.9743E-07 &2.4756E-07 &2.8771E-07 & 9.1714E-08 \\
$^{64}$Zn&4.1170E-06 &5.4031E-06 &6.3524E-06 &5.5728E-06 &1.0068E-05 & 1.2144E-06 \\
$^{66}$Zn&5.3276E-06 &7.1634E-06 &8.5197E-06 &6.8650E-06 &6.0776E-06 & 5.0221E-06 \\
$^{67}$Zn&2.1888E-12 &2.7503E-12 &3.2022E-12 &1.0714E-12 &2.5332E-12 & 2.9738E-12 \\
$^{68}$Zn&2.4497E-09 &3.4762E-09 &4.2682E-09 &2.9172E-09 &2.4247E-09 & 6.0432E-09 \\
$^{70}$Zn&1.6091E-19 &2.1266E-19 &2.3458E-19 &1.4498E-23 &2.0416E-19 & 2.5513E-19 \\
\hline \hline
\end{longtable}

\begin{longtable}{c c c c c c c }
\caption{\label{radio_tab}
Synthesized mass $(M_{\odot})$ of radioactive species in
  different models}\\
\hline \hline
Species & \multicolumn{6}{c}{model} \\
 & \emph{2\_1\_2} & \emph{2\_2\_2} & \emph{2\_3\_2} & \emph{1\_2\_2} &
  \emph{2\_2\_1} & \emph{2\_2\_3} \\
\endhead
\hline
$^{22}$Na&5.2892E-08 &9.2173E-08 &1.0290E-07 &8.0645E-08 &9.9089E-08 & 4.1895E-08 \\
$^{26}$Al&1.3078E-06 &1.1027E-06 &8.0954E-07 &1.0024E-06 &4.7110E-07 & 2.8228E-06 \\
$^{36}$Cl&7.6022E-07 &7.2749E-07 &6.8645E-07 &8.1239E-07 &2.8832E-07 & 1.1296E-06 \\
$^{39}$Ar&4.3326E-09 &2.6531E-09 &1.7073E-09 &2.5865E-09 &5.3153E-10 & 1.3045E-08 \\
$^{40}$K &1.5893E-08 &1.1857E-08 &8.8932E-09 &1.2172E-08 &3.5607E-09 & 1.5243E-08 \\
$^{41}$Ca&3.3488E-06 &1.6224E-06 &6.4171E-07 &1.8647E-06 &1.0311E-06 & 2.3714E-06 \\
$^{44}$Ti&2.7207E-06 &2.8370E-06 &3.2374E-06 &2.9291E-06 &3.3876E-06 & 1.3530E-06 \\
$^{48}$V &2.7174E-08 &1.6707E-08 &1.0033E-08 &1.9273E-08 &1.0073E-08 & 4.0923E-08 \\
$^{49}$V &6.0178E-08 &5.0885E-08 &5.9430E-08 &3.3875E-08 &3.0780E-08 & 2.8737E-07 \\
$^{53}$Mn&2.0549E-04 &1.9908E-04 &1.9552E-04 &1.6387E-05 &1.9147E-04 & 2.7557E-04 \\
$^{55}$Fe&2.0833E-03 &2.0297E-03 &1.9799E-03 &2.1947E-04 &1.9797E-03 & 2.3335E-03 \\
$^{59}$Fe&8.7600E-13 &9.1399E-13 &9.4831E-13 &8.2378E-17 &8.7169E-13 & 1.1308E-12 \\
$^{60}$Fe&4.7700E-15 &5.2457E-15 &5.5233E-15 &1.2764E-19 &4.9474E-15 & 6.8064E-15 \\
$^{56}$Co&1.0058E-04 &9.8004E-05 &9.3479E-05 &4.5078E-05 &9.2616E-05 & 1.2481E-04 \\
$^{57}$Co&1.0771E-03 &1.0724E-03 &1.0677E-03 &1.1285E-04 &1.0531E-03 & 1.1640E-03 \\
$^{60}$Co&2.5561E-10 &2.7429E-10 &2.8596E-10 &1.8310E-13 &2.6598E-10 & 3.1485E-10 \\
$^{56}$Ni&2.9992E-01 &3.0461E-01 &3.0648E-01 &2.8599E-01 &3.1625E-01 & 2.5917E-01 \\
$^{57}$Ni&1.1578E-02 &1.2281E-02 &1.2648E-02 &1.0643E-02 &1.1183E-02 & 1.4214E-02 \\
$^{59}$Ni&4.6795E-04 &4.7814E-04 &4.8911E-04 &8.0744E-05 &4.6775E-04 & 5.2852E-04 \\
$^{60}$Ni&5.1163E-03 &5.2800E-03 &5.4430E-03 &2.3015E-04 &5.2058E-03 & 5.6263E-03 \\
$^{61}$Ni&2.8036E-06 &2.9852E-06 &3.1293E-06 &2.4920E-08 &2.9360E-06 & 3.2136E-06 \\
$^{62}$Ni&5.3714E-06 &5.7961E-06 &6.1280E-06 &8.1844E-09 &5.6628E-06 & 6.4208E-06 \\
$^{63}$Ni&2.2585E-11 &2.5002E-11 &2.6783E-11 &3.4056E-15 &2.4055E-11 & 2.9699E-11 \\
\hline\hline
\end{longtable}

\end{document}